\documentclass[english,twocolumn,prb,showpacs,preprintnumbers,amsmath,amssymb]{revtex4}
\usepackage[T1]{fontenc}
\usepackage[latin1]{inputenc}
\usepackage{longtable}
\usepackage{booktabs}
\usepackage{amsmath}
\usepackage{graphicx}
\usepackage{amssymb}

\makeatletter

\providecommand{\tabularnewline}{\\}

\@ifundefined{textcolor}{}
{%
 \definecolor{BLACK}{gray}{0}
 \definecolor{WHITE}{gray}{1}
 \definecolor{RED}{rgb}{1,0,0}
 \definecolor{GREEN}{rgb}{0,1,0}
 \definecolor{BLUE}{rgb}{0,0,1}
 \definecolor{CYAN}{cmyk}{1,0,0,0}
 \definecolor{MAGENTA}{cmyk}{0,1,0,0}
 \definecolor{YELLOW}{cmyk}{0,0,1,0}
 }

\usepackage{epsfig}

\usepackage{dcolumn}
\usepackage{bm}
\usepackage{wasysym}

\usepackage{marvosym}

\usepackage{color}

\makeatletter

\usepackage{ulem}

\makeatother

\makeatother

\usepackage{babel}

\begin{document}

\title{Electronic properties of a biased graphene bilayer}

\author{Eduardo V. Castro$^{1,2}$, K. S. Novoselov$^{3}$, S. V. Morozov$^{3}$,
N.~M.~R.~Peres$^{4}$, J.~M.~B.~Lopes~dos~Santos$^{1}$, Johan
Nilsson$^{5}$, F. Guinea$^{2}$, A. K. Geim$^{3}$, and A. H. Castro
Neto$^{5}$}

\affiliation{$^{1}$CFP and Departamento de F\'{\i}sica, Faculdade de Ciências
Universidade do Porto, P-4169-007 Porto, Portugal}

\affiliation{$^{2}$Instituto de Ciencia de Materiales de Madrid, CSIC, Cantoblanco, E-28049 Madrid, Spain}

\affiliation{$^{3}$Department of Physics and Astronomy, University of Manchester,
Manchester, M13~9PL, UK}

\affiliation{$^{4}$Center of Physics and Departamento de F\'{\i}sica, Universidade
do Minho, P-4710-057 Braga, Portugal}

\affiliation{$^{5}$Department of Physics, Boston University, 590 Commonwealth
Avenue, Boston, MA 02215,USA}

\date{\today{}}
\begin{abstract}
We study, within the tight-binding approximation, the electronic properties
of a graphene bilayer in the presence of an external electric field
applied perpendicular to the system -- \emph{biased bilayer}. The
effect of the perpendicular electric field is included through a parallel
plate capacitor model, with screening correction at the Hartree level.
The full tight-binding description is compared with its 4-band and
2-band continuum approximations, and the 4-band model is shown to
be always a suitable approximation for the conditions realized in
experiments. The model is applied to real biased bilayer devices,
either made out of SiC or exfoliated graphene, and good agreement
with experimental results is found, indicating that the model is capturing
the key ingredients, and that a finite gap is effectively being controlled
externally. Analysis of experimental results regarding the electrical
noise and cyclotron resonance further suggests that the model can
be seen as a good starting point to understand the electronic properties
of graphene bilayer. Also, we study the effect of electron-hole asymmetry
terms, as the second-nearest-neighbor hopping energies $t'$ (in-plane)
and $\gamma_{4}$ (inter-layer), and the on-site energy $\Delta$.
\end{abstract}


\maketitle

%

\section{Introduction}

\label{sec:BilayerIntro}

The double layer graphene system -- the so-called bilayer graphene
(BLG) -- is now a subject of considerable interest due to its unusual
properties,\citep{MAFssc07,MAFepj07,AberF07,GNPssc} dissimilar in
large extent to those of the single layer graphene (SLG).\citep{NGPrmp}
The integer quantum Hall effect (QHE) is a paradigmatic case; characterized
by the absence of a plateau at the Dirac point,\citep{NMcCM+06} and
thus still anomalous, it is associated with massive Dirac fermions
and two zero energy modes.\citep{MF06} 

One of the most remarkable properties of BLG is the ability to open
a gap in the spectrum by electric field effect~--~biased BLG.\citep{MF06}
This has been shown both experimentally and theoretically, providing
the first semiconductor with externally tunable gap.\citep{OBS+06,OHL+07,CNM+06,GNP06,McC06,MSB+06,AA07,Guinea07,MLS+09,ZTG+09}
In the absence of external perpendicular electric field~--~unbiased
BLG~--~the system is characterized by four bands, two of them touching
each other parabolically at zero energy, and giving rise to the massive
Dirac fermions mentioned above, and other two separated by an energy
$\pm t_{\perp}$. Hence, an unbiased BLG is a two-dimensional zero-gap
semiconductor.\citep{MF06,NMcCM+06,NNGP06} At the neutrality point
the conductivity shows a minimum of the order of the conductance quantum,\citep{NMcCM+06,MNK+07,KA06,Kat06a,Kat06b,Cserti06,NNGP06,CCD07,SB07}
a property shared with SLG.\citep{KNssc07} This prevents standard
device applications where the presence of a finite gap producing high
on-off current ratios is of paramount importance. The fact that a
simple perpendicular electric field is enough to open a gap, and even
more remarkable, to control its size, clearly demonstrates the potential
of this system for carbon-based electronics.\citep{NNG+06,PVP07}

The biased BLG reveals interesting properties on its own. The gap
has shown to be robust in the presence of disorder,\citep{NN06,CPS06,NNG+07}
induced either by impurities or dilution, but is completely absent
in rotated (non $AB$-stacked) bilayers, where the SLG linear dispersion
is recovered.\citep{dSPN,HVM+08} The band structure near the gap
shows a ''Mexican-hat'' like behavior, with a low doping Fermi surface
which is a ring.\citep{GNP06} Such a topologically nontrivial Fermi
surface leads to an enhancement of electron-electron interactions,
and to a ferromagnetic instability at low enough density of carriers.\citep{Sta07,CPS+07}
In the presence of a perpendicular magnetic field, the biased BLG
shows cyclotron mass renormalization and an extra plateau at zero
Hall conductivity, signaling the presence of a sizable gap at the
neutrality point.\citep{McC06,CNM+06,PPV07} Gaps can also be induced
in stacks with more than two layers as long as the stacking order
is of the rhombohedral-type,\citep{GNP06,AA07} although screening
effects may become important in doped systems with increasing number
of layers.\citep{Guinea07} Recently, a ferromagnetic proximity effect
was proposed as a different mechanism which can also open a gap in
the spectrum of the BLG, leading to a sizable magnetoresistive effect.\citep{SKZ08}
Strain applied to the biased BLG has also shown to produce further
gap modulation.\citep{RK09}

In this paper the electronic properties of a biased BLG are studied
within a full tight-binding model, which enables the analysis of the
whole bandwidth, validating previous results obtained using low-energy
effective models. The screening of the applied perpendicular electric
field is obtained within a self-consistent Hartree approach, and a
comparison with experiments is provided. The effect of the bias in
the cyclotron mass and cyclotron resonance is addressed, and the results
are shown to agree well with experimental measurements.

The paper is organized as follows: in Sec.~\ref{sec:BilayerModel}
the lattice structure of BLG and the tight-binding Hamiltonian are
presented; bulk electronic properties are discussed in Sec~\ref{sec:bulkelecprop},
with particular emphasis on the screening correction; the effect of
a perpendicular magnetic field is studied in Sec.~\ref{sec:BilayerMFE};
Sec.~\ref{sec:conclusions} contains our conclusions. We have also
included three appendices: Appendix~\ref{secap:Dn} provides details
on the calculation of the density asymmetry between layers for a finite
bias; in Appendix~\ref{secap:BilayerDOS} we give the analytical
expression for the biased BLG density of states, valid over the entire
energy spectrum; analytical expressions for the cyclotron mass obtained
within the full tight-binding model are given in Appendix~\ref{secap:Bilayermc}.

%

\section{Model}

\label{sec:BilayerModel} 

Here we consider only $AB$-Bernal stacking, where the top layer has
its $A$ sublattice on top of sublattice $B$ of the bottom layer.
We use indices~1 and~2 to label the top and bottom layer, respectively.
The unit cell of a bilayer has twice the number of atoms of a single
layer. The basis vectors may be written as $\mathbf{a}_{1}=a\,\textrm{ê}_{x}$
and $\mathbf{a}_{2}=a(\textrm{ê}_{x}-\sqrt{3}\,\textrm{ê}_{y})/2$,
where $a=2.46\,\textrm{Å}$.

In the tight-binding approximation, the in-plane hopping energy, $t$,
and the inter-layer hopping energy, $t_{\perp}$, define the most
relevant energy scales. The simplest tight-binging Hamiltonian describing
non-interacting $\pi-$electrons in BLG reads:\begin{equation}
H_{TB}=\sum_{i=1}^{2}H_{i}+t_{\perp}\sum_{\mathbf{R},\sigma}\big[a_{1,\sigma}^{\dagger}(\mathbf{R})b_{2,\sigma}(\mathbf{R})+\textrm{h.c.}\big]+H_{V},\label{eq:Hbilayer}\end{equation}
with the SLG Hamiltonian\begin{multline}
H_{i}=-t\sum_{\mathbf{R},\sigma}\big[a_{i,\sigma}^{\dagger}(\mathbf{R})b_{i,\sigma}(\mathbf{R})+a_{i,\sigma}^{\dagger}(\mathbf{R})b_{i,\sigma}(\mathbf{R}-\mathbf{a}_{1})\\
+a_{i,\sigma}^{\dagger}(\mathbf{R})b_{i,\sigma}(\mathbf{R}-\mathbf{a}_{2})+\textrm{h.c.}\big],\label{eq:Hslg}\end{multline}
where $a_{i,\sigma}(\mathbf{R})$ {[}$b_{i,\sigma}(\mathbf{R})${]}
is the annihilation operator for electrons at position $\mathbf{R}$
in sublattice $Ai$ ($Bi$), $i=1,2$, and spin $\sigma$. The in-plane
hopping $t$ can be inferred from the Fermi velocity in graphene $v_{\textrm{F}}=ta\hbar^{-1}\sqrt{3}/2\approx10^{6}\,\mbox{ms}^{-1}$,\citep{NGM+05}
yielding $t\approx3.1\,\mbox{eV}$, in good agreement with what is
found experimentally for graphite.\citep{TDD77} This value also agrees
with a recent Raman scattering study of the electronic structure of
BLG.\citep{MNE+07} As regards the inter-layer hopping $t_{\perp}$,
angle-resolved photoemission spectroscopy (ARPES) measurements in
epitaxial BLG give $t_{\perp}\approx0.43\,\mbox{eV}$,\citep{OBS+06}
and Raman scattering for BLG obtained by micro-mechanical cleavage
of graphite yields $t_{\perp}\approx0.30\,\mbox{eV}$.\citep{MNE+07}
The experimental value for bulk graphite is $t_{\perp}\approx0.39\,\mbox{eV}$,\citep{MMD79}
which means that for practical purposes we can always assume $t_{\perp}/t\sim0.1\ll1$.
This values for $t$ and $t_{\perp}$ compare fairly well with what
is obtained from fist-principles calculations for graphite\citep{CGM91}
using the well established Slonczewski-Weiss-McClure (SWM) parametrization
model\citep{McClure57,SW58} to fit the bands near the Fermi energy.
The SWM model assumes extra parameters that can also be incorporated
in a tight-binding model for BLG. Namely, the inter-layer second-NN
hoppings $\gamma_{3}$ and $\gamma_{4}$, where $\gamma_{3}$ connects
different sublattices ($B1-A2$) and $\gamma_{4}$ equal sublattices
($A1-A2$ and $B1-B2$). Additionally, there is an on-site energy
$\Delta$ reflecting the inequivalence between sublattices $A1,B2$
and $B1,A2$ -- the former project exactly on top of each other while
the latter lay on the hexagon center of the other layer. The consequences
of these extra terms for the band structure obtained from Eq.~(\ref{eq:Hbilayer})
are well known: $\gamma_{3}$ induces trigonal warping and both $\gamma_{4}$
and $\Delta$ give rise to electron-hole asymmetry.\citep{PGN06,MF06,PP06}
The in-plane second-NN hopping energy $t'$ is not considered in the
usual tight-binding parametrization of the SWM model. Nevertheless,
this term can have important consequences since it breaks particle-hole
symmetry but does not modify the Dirac spectrum. Typical values are
given in Table~\ref{tab:params} as obtained in recent experiments,
except for $t'$ quoted from density functional theory (DFT) calculations.

\begin{table}

\begin{centering}
\begin{longtable}{cccc}
\hline
\toprule 
~~~$\gamma_{3}/t$\citep{MNE+07,KCM+09}~~~ & ~~~$\gamma_{4}/t$\citep{MNE+07,ZLB+08,KCM+09}~~~ & ~~~$\Delta/t$\citep{ZLB+08,LHJ+09,KCM+09}~~~ & ~~~$t'/t$\citep{RMT+02}~~~\tabularnewline
\hline
\endhead
$0.03-0.1$ & $0.04-0.07$ & $0.005-0.008$ & $\sim0.04$\tabularnewline
\bottomrule
\end{longtable}
\par\end{centering}

\caption{\label{tab:params}Approximate parameter values as obtained in recent
experiments (except for $t'$ quoted from DFT calculations).}

\end{table}

We are interested in the properties of BLG in the presence of a perpendicular
electric field~--~the biased BLG. The effect of the induced energy
difference between layers, parametrized by $V$, may be accounted
for by adding $H_{V}$ to Eq.~(\ref{eq:Hbilayer}), with $H_{V}$
given by\begin{equation}
H_{V}=\frac{V}{2}\sum_{\mathbf{R},\sigma}\big[n_{A1}(\mathbf{R})+n_{B1}(\mathbf{R})-n_{A2}(\mathbf{R})-n_{B2}(\mathbf{R})\big],\label{eq:HV}\end{equation}
 where $n_{Ai}(\mathbf{R})$ and $n_{Bi}(\mathbf{R})$ are number
operators.

%

\section{Bulk electronic properties}

\label{sec:bulkelecprop}

Introducing the Fourier components $a_{i,\sigma,\mathbf{k}}$ and
$b_{i,\sigma,\mathbf{k}}$ of operators $a_{i,\sigma}(\mathbf{R})$
and $b_{i,\sigma}(\mathbf{R})$, respectively, with the layer index
$i=1,2$, we can rewrite Eq.~(\ref{eq:Hbilayer}) as $H=\sum_{\mathbf{k},\sigma}\psi_{\sigma,\mathbf{k}}^{\dagger}H_{\mathbf{k}}\psi_{\sigma,\mathbf{k}}$,
where $\psi_{\sigma,\mathbf{k}}^{\dagger}=[a_{1,\sigma,\mathbf{k}}^{\dagger},b_{1,\sigma,\mathbf{k}}^{\dagger},a_{2,\sigma,\mathbf{k}}^{\dagger},b_{2,\sigma,\mathbf{k}}^{\dagger}]$
is a four component spinor, and $H_{\mathbf{k}}$ is given by\begin{equation}
H_{\mathbf{k}}=\left(\begin{array}{cccc}
V/2 & -ts_{\mathbf{k}} & 0 & -t_{\perp}\\
-ts_{\mathbf{k}}^{*} & V/2 & 0 & 0\\
0 & 0 & -V/2 & -ts_{\mathbf{k}}\\
-t_{\perp} & 0 & -ts_{\mathbf{k}}^{*} & -V/2\end{array}\right).\label{eq:HkBilayer}\end{equation}
The factor $s_{\mathbf{k}}=1+e^{i\mathbf{k\cdot}\mathbf{a}_{1}}+e^{i\mathbf{k\cdot}\mathbf{a}_{2}}$
determines the matrix elements for the SLG Hamiltonian in reciprocal
space ($t_{\perp}=0$, $V=0$), from which the SLG dispersion is obtained,
$\epsilon_{\mathbf{k}}=\pm t|s_{\mathbf{k}}|$. The resultant conduction
($+$) and valence ($-$) bands touch each other in a conical way
at the corners of the first Brillouin zone (BZ), the $K$ and $K'$
points.\citep{NGPrmp} This touching occurs at zero energy, the Fermi
energy for undoped graphene. The 4--band continuum approximation for
Eq.~(\ref{eq:HkBilayer}), valid at energy scales $E\ll t$, may
be obtained by introducing the small wave vector $\mathbf{q}$ which
measures the difference between $\mathbf{k}$ and the corners of the
BZ. Linearizing the factor $s_{\mathbf{k}}$ around the $K$ points
Eq.~(\ref{eq:HkBilayer}) reads\begin{equation}
H_{K}=\left(\begin{array}{cccc}
V/2 & v_{\textrm{F}}pe^{-i\varphi_{\mathbf{p}}} & 0 & -t_{\perp}\\
v_{\textrm{F}}pe^{i\varphi_{\mathbf{p}}} & V/2 & 0 & 0\\
0 & 0 & -V/2 & v_{\textrm{F}}pe^{-i\varphi_{\mathbf{p}}}\\
-t_{\perp} & 0 & v_{\textrm{F}}pe^{i\varphi_{\mathbf{p}}} & -V/2\end{array}\right),\label{eq:HKbl}\end{equation}
where $\mathbf{p}=\hbar\mathbf{q}$ and $\varphi_{\mathbf{p}}=\tan^{-1}(p_{y}/p_{x})$.
Around the $K'$ points Eq.~(\ref{eq:HKbl}) with complex conjugate
matrix elements defines $H_{K'}$.\citep{NNP+05,MF06}

Equation~(\ref{eq:HKbl}) can be further simplified if one assumes
$v_{\textrm{F}}p,V\ll t_{\perp}$. By eliminating high energy states
perturbatively we can write a two-band effective Hamiltonian describing
low-energy states whose electronic amplitude is mostly localized on
$B1$ and $A2$ sites. Near the $K$ points the resulting Hamiltonian
may be written as\begin{equation}
H_{eff}=-\left(\begin{array}{cc}
-V/2 & e^{i2\varphi_{\mathbf{p}}}v_{\textrm{F}}^{2}p^{2}/t_{\perp}\\
e^{-i2\varphi_{\mathbf{p}}}v_{\textrm{F}}^{2}p^{2}/t_{\perp} & V/2\end{array}\right),\label{eq:Heff}\end{equation}
whereas the complex conjugate matrix elements should be taken for
a low-energy description around the $K'$ points. The two-component
wave functions have the form $\Phi=(\phi_{B1},\phi_{A2})$.\citep{NNP+05,MF06,MGV07}

In the following we discuss the electronic structure resulting from
the tight-binding Hamiltonian~(\ref{eq:HkBilayer}), and comment
on the approximations given above by Eqs.~(\ref{eq:HKbl}) and~(\ref{eq:Heff}).


\subsection{Electronic structure}

\label{subsec:electStruct} 

Let us briefly discuss the electronic structure of the biased BLG
using the full tight-binding Hamiltonian given by Eq.~(\ref{eq:Hbilayer}).
The spectrum of Eq.~(\ref{eq:Hbilayer}) for $V\neq0$ reads: \begin{equation}
E_{\mathbf{k}}^{\pm\pm}(V)=\pm\sqrt{\epsilon_{\mathbf{k}}^{2}+\frac{t_{\perp}^{2}}{2}+\frac{V^{2}}{4}\pm\sqrt{t_{\perp}^{4}/4+(t_{\perp}^{2}+V^{2})\epsilon_{\mathbf{k}}^{2}}}.\label{eq:Ekbias}\end{equation}
As can be seen from Eq.~(\ref{eq:Ekbias}), the $V=0$ gapless system
turns into a semiconductor with a gap controlled by $V$. Moreover,
the two bands close to zero energy are deformed near the corners of
the BZ,\citealp{NGPrmp} so that the minimum of $|E_{\mathbf{k}}^{\pm-}(V)|$
no longer occurs at these corners. As a consequence, the low doping
Fermi surface is completely different from the $V=0$ case, with its
shape controlled by $V$.

\begin{figure}
\begin{centering}
\includegraphics[width=0.98\columnwidth]{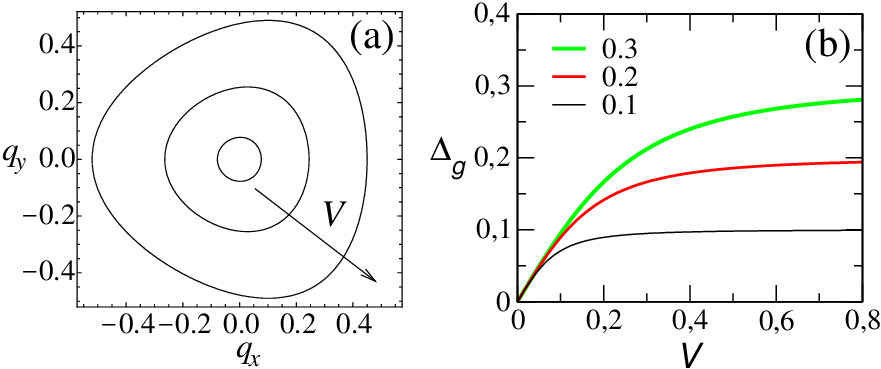}
\par\end{centering}

\caption{\label{cap:bandstructV}(Color online) (a)~Solution of Eq.~(\ref{eq:minima})
for $V=t_{\perp}/2,2t_{\perp},4t_{\perp}$. (b)~$\Delta_{g}$ vs
$V$ for various $t_{\perp}$ values. Energy is given in units of
$t$ and momentum in units of $a^{-1}$.}

\end{figure}

It can be readily shown that the minimum of sub-band $E_{\mathbf{k}}^{+-}(V)$
{[}or equivalently, the maximum of $E_{\mathbf{k}}^{--}(V)${]} occurs
for all $\mathbf{k}$'s satisfying \begin{equation}
\epsilon_{\mathbf{k}}^{2}=\alpha(V,t_{\perp})\,,\label{eq:minima}\end{equation}
with $\alpha(V,t_{\perp})=(V^{4}/4+t_{\perp}^{2}V^{2}/2)/(V^{2}+t_{\perp}^{2})$
-- note that $\partial E_{\mathbf{k}}^{\pm-}/\partial\epsilon_{\mathbf{k}}=0$
at the desired extrema. Equation~(\ref{eq:minima}) has solutions
for $\sqrt{\alpha}\leq3t$ ($3t$ is half of the single layer bandwidth).
When $\sqrt{\alpha}>3t$ the minimum of $E_{\mathbf{k}}^{+-}(V)$
occurs at the $\Gamma$ point. Figure~\ref{cap:bandstructV}(a) shows
the solution of Eq.~(\ref{eq:minima}) around the $K$ point for
$V=t_{\perp}/2,2t_{\perp},4t_{\perp}$ (around the $K'$ point the
figure is rotated by $\pi/3$). At low doping the Fermi sea acquires
a line shape given by the solution of Eq.~(\ref{eq:minima}), the
line width being determined by the doping level. As can be seen in
Fig.~\ref{cap:bandstructV}(a), when $V<t_{\perp}$ the Fermi sea
approaches a ring, the Fermi ring, centered at the BZ corners.\citep{GNP06,Sta07}
As $V$ is increased there is an apparent trigonal distortion showing
up, which originates from the single layer dispersion in Eq.~(\ref{eq:minima}).

The existence of a Fermi ring is easily understood using the continuum
version of Eq.~(\ref{eq:Ekbias}), i.e., the eigenvalues of Eq.~(\ref{eq:HKbl}).
This amounts to substituting the single layer dispersion in Eq.~(\ref{eq:Ekbias})
by $v_{\textrm{F}}p$, which immediately implies cylindrical symmetry
around $K$ and $K'$. If we further assume that $v_{\textrm{F}}p\ll V\ll t_{\perp}$
holds, Eq.~(\ref{eq:Ekbias}) is then well approximated by the {}``Mexican
hat'' dispersion,\citep{GNP06} \begin{equation}
E^{\pm-}(V)\approx\pm\frac{V}{2}\mp\frac{Vv_{\textrm{F}}^{2}}{t_{\perp}^{2}}p^{2}\pm\frac{v_{\textrm{F}}^{4}}{t_{\perp}^{2}V}p^{4},\label{eq:mexhat}\end{equation}
which explains the Fermi ring. If, instead, we have $V<v_{\textrm{F}}p\ll t_{\perp}$,
we can approximate Eq.~(\ref{eq:Ekbias}) by\begin{equation}
E^{\pm-}(V)\approx\pm\sqrt{V^{2}/4+v_{\textrm{F}}^{4}p^{4}/t_{\perp}^{2}},\label{eq:effEkbias}\end{equation}
which corresponds exactly to the eigenvalues of the effective two-band
Hamiltonian in Eq.~(\ref{eq:Heff}). Note that no continuum approximation
can produce the trigonal distortion shown in Fig.~\ref{cap:bandstructV}(a).

The gap between conduction and valence bands, $\Delta_{g}$, is twice
the minimum value of $E_{\mathbf{k}}^{+-}(V)$ due to electron-hole
symmetry, and is given by,\begin{equation}
\Delta_{g}=\begin{cases}
\sqrt{t_{\perp}^{2}V^{2}/(t_{\perp}^{2}+V^{2})} & V\leq V_{\textrm{c}}\\
2t\sqrt{9+\frac{t_{\perp}^{2}}{2t^{2}}+\frac{V^{2}}{4t^{2}}-\sqrt{\frac{t_{\perp}^{4}}{4t^{4}}+9\frac{t_{\perp}^{2}+V^{2}}{t^{2}}}} & V>V_{\textrm{c}}\end{cases},\label{eq:gapV}\end{equation}
where $V_{\textrm{c}}=[18t^{2}-t_{\perp}^{2}+(18^{2}t^{4}+t_{\perp}^{4})^{1/2}]^{1/2}\simeq6t$,
the approximation being valid for $t_{\perp}\ll t$. From Eq.~(\ref{eq:gapV})
it can be seen that for both $V\ll t_{\perp}$ and $V\gg t$ one finds
$\Delta_{g}\sim V$. However, there is a region for $t_{\perp}\lesssim V\lesssim6t$
where the gap shows a plateau $\Delta_{g}\sim t_{\perp}$, as depicted
in Fig.~\ref{cap:bandstructV}(b). The plateau ends when $V\simeq6t$
(not shown). 


\subsection{Screening of the external field}

\label{subsec:screen}

So far we have considered $V$, i.e. the electrostatic energy difference
between layers felt by a single electron, as a band parameter that
controls the gap. However, the parameter $V$ can be related with
the perpendicular electric field applied to BLG, avoiding the introduction
of an extra free parameter in the present theory.

Let us call $\mathbf{E}=E\hat{e}_{z}$ the perpendicular electric
field felt by electrons in BLG. The corresponding electrostatic energy
$U(z)$ for an electron of charge $-e$ is related to the electric
field as $eE=\partial U(z)/\partial z$, and thus $V$ is given by\begin{equation}
V=U(z_{1})-U(z_{2})=eEd,\label{eq:VofE}\end{equation}
where $z_{1}$ and $z_{2}$ are the positions of layer~1 and~2,
respectively, and $d\equiv z_{1}-z_{2}=3.4\,\mbox{Å}$ is the inter-layer
distance. Given the experimental conditions, the value of $E$ can
be calculated under a few assumptions, as detailed in the following. 


\subsubsection{External field in real systems}

\label{subsubsec:screen1}

If we assume the electric field $E$ in Eq.~\eqref{eq:VofE} to be
due exclusively to the external electric field applied to BLG, $E=E_{ext}$,
all we need in order to know $V$ is the value of $E_{ext}$,\begin{equation}
V=eE_{ext}d.\label{eq:VofEext}\end{equation}
The experimental realization of a biased BLG has been achieved in
epitaxial BLG through chemical doping\citep{OBS+06,ZSF+08a} and in
back gated exfoliated BLG.\citep{CNM+06,OHL+07} In either case the
value of $E_{ext}$ can be extracted assuming a simple parallel plate
capacitor model.

\begin{figure}
\noindent \begin{centering}
\includegraphics[width=0.98\columnwidth]{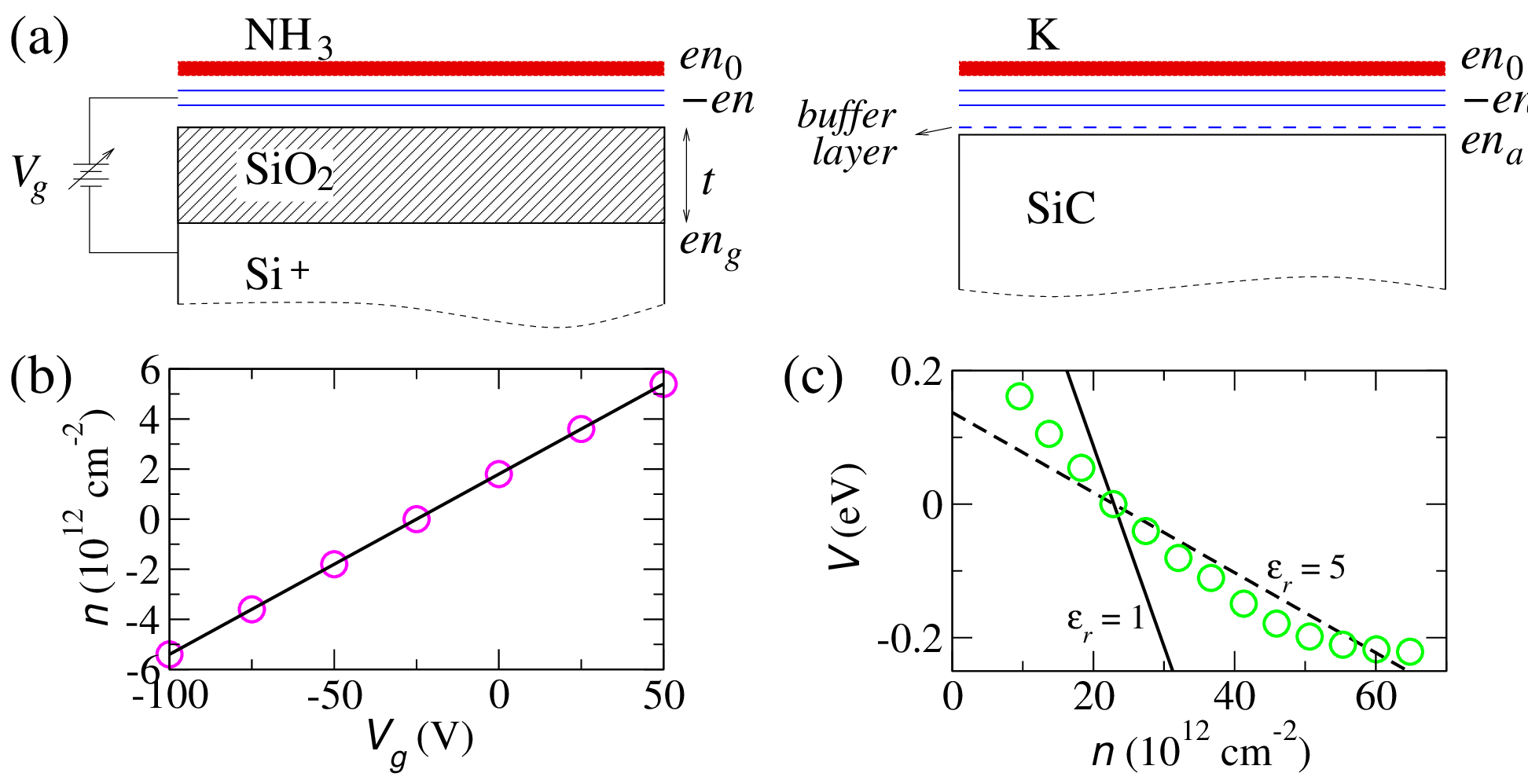}
\par\end{centering}

\caption{\label{fig:device}(Color online) (a)~Biased BLG devices. (b)~$n$
vs $V_{g}$ for the left device shown in~(a): experimental data is
shown as symbols;\citep{CNM+06} the line is a linear fit with Eq.~\eqref{eq:nVg}.
(c)~$V$ vs $n$ for the right BLG device shown in~(a): symbols
are experimental data from Ref.~\onlinecite{OBS+06}; the lines are
the result of Eq.~\eqref{eq:VnUnSohta}.}

\end{figure}

In the case of exfoliated BLG, devices are prepared by micromechanical
cleavage of graphite on top of an oxidized silicon wafer ($300\,\mbox{nm}$
of $\mbox{SiO}_{2}$), as shown in the left panel of Fig.~\ref{fig:device}(a).
A back gate voltage $V_{g}$ applied between the sample and the Si
wafer induces charge carriers due to the electric field effect, resulting
in carrier densities $n_{g}=\beta V_{g}$ relatively to half-filling
($n_{g}>0$ for electrons and $n_{g}<0$ for holes). The geometry
of the resulting capacitor determines the coefficient $\beta$. In
particular, the electric field inside the oxidized layer is $E_{ox}=en_{g}/(\varepsilon_{\textrm{SiO}_{2}}\varepsilon_{0})$,
where $\varepsilon_{\textrm{SiO}_{2}}$ and $\varepsilon_{0}$ are
the permittivities of $\mbox{SiO}_{2}$ and free space, respectively.
This implies a gate voltage $V_{g}=en_{g}t/(\varepsilon_{\textrm{SiO}_{2}}\varepsilon_{0})$,
from which we obtain the coefficient $\beta=\varepsilon_{\textrm{SiO}_{2}}\varepsilon_{0}/(et)$.
For a $\mbox{SiO}_{2}$ thickness $t=300\,\mbox{nm}$ and a dielectric
constant $\varepsilon_{\textrm{SiO}_{2}}=3.9$ we obtain $\beta\cong7.2\times10^{10}\,\textrm{cm}^{-2}/\textrm{V}$,
which is in agreement with the values found experimentally.\citep{NGM+05,ZTS+05,NMcCM+06}
In order to control independently the gap value and the Fermi level,
in Ref.~\onlinecite{CNM+06} the devices have been chemically doped
by deposition of NH$_{3}$ on top of the upper layer, which adsorbed
on graphene and effectively acted as a top gate providing a fixed
electron density $n_{0}$.\citep{SGM+07} Charge conservation then
implies a total density $n$ in BLG given by $n=n_{g}+n_{0}$, or
in terms of the applied gate voltage,\begin{equation}
n=\beta V_{g}+n_{0}.\label{eq:nVg}\end{equation}
In Fig.~\ref{fig:device}(b) the charge density in BLG is shown as
a function of $V_{g}$. The symbols are the experimental result obtained
from Hall effect measurements,\citep{CNM+06} and the line is a linear
fit with Eq.~\eqref{eq:nVg}. The fit provides $n_{0}$, which for
this particular experimental realization is $n_{0}\simeq1.8\times10^{12}\,\mbox{cm}^{-2}$,
and validates the parallel plate capacitor model applied to the back
gate, since the fitted $\beta\simeq7.2\times10^{10}\,\textrm{cm}^{-2}/\textrm{V}$
is in excellent agreement with the theoretical value. Extending the
parallel plate capacitor model to include the effect of dopants, the
external field $E_{ext}$ is the result of charged surfaces placed
above and below BLG. The accumulation or depletion layer in the Si
wafer contributes with an electric field $E_{\textrm{b}}=en_{g}/(2\varepsilon_{r}\varepsilon_{0})$,
while dopants above BLG effectively provide the second charged surface
with electric field $E_{\textrm{t}}=-en_{0}/(2\varepsilon_{r}\varepsilon_{0})$.
A relative dielectric constant $\varepsilon_{r}$ different from unity
may be attributed to the presence of SiO$_{2}$ below and vacuum on
top, which gives  $\varepsilon_{r}\approx(\varepsilon_{\textrm{SiO}_{2}}+1)/2\approx2.5$,
a value that can be slightly different due to adsorption of water
molecules.\citep{SGM+07,MVJ+08} Adding the two contributions, $E_{ext}=E_{\textrm{b}}+E_{\textrm{t}}$,
and making use of the charge conservation relation, we arrive at an
electrostatic energy difference $V$ {[}Eq.~\eqref{eq:VofEext}{]}
that depends linearly on the BLG density, \begin{equation}
V=\left(\frac{n}{n_{0}}-2\right)\frac{e^{2}n_{0}d}{2\varepsilon_{r}\varepsilon_{0}}.\label{eq:VnUnSgeim}\end{equation}
In treating the dopants as a homogeneous charged layer we ignore possible
lattice distortion induced by adsorbed molecules, as well as the electric
field due to the NH$_{3}$ electric dipole, which may contribute to
the gap in the spectrum. However, it has been shown recently\citep{RPC+08}
that for NH$_{3}$ these effects counteract, giving rise to a much
smaller gap than other dopant molecules,\citep{BK08} as for instance
NH$_{2}$ and CH$_{3}$. For the biased BLG realized in Ref.~\onlinecite{OHL+07},
independence of Fermi level and carrier density was achieved with
a real top gate, which makes the parallel plate capacitor model a
suitable approximation in that case.

In the case of epitaxial BLG, devices are grown on SiC by thermal
decomposition of the Si-face.\citep{foot1} The substrate is fixed
(SiC), and graphene behavior develops for carbon layers above the
buffer layer,\citep{VFH+07,MP07,KIC+07,VMV+08} as schematically shown
in the right panel of Fig.~\ref{fig:device}(a).\citep{foot2} Due
to charge transfer from substrate to film, the as-prepared BLG devices
appear electron doped with density $n_{\textrm{a}}$. First-principles
calculations indicate that such doping is coming from interface states
that develop between the buffer layer and the Si-terminated substrate.\citep{VFH+07,MP07}
{[}Scanning tunneling microscopy (STM) measurements corroborate the
presence of interface states.\citep{MVN+07,RGC+07,BZY+07,VMV+08}{]}
From the point of view of our theoretical approach, we may interpret
these interface states as an effective depletion layer that provides
the external electric field necessary to make the system a biased
BLG. In Ref.~\onlinecite{OBS+06} the BLG density $n$ was varied
by doping the system with potassium (K) on top of the upper layer
{[}see Fig.~\ref{fig:device}(a){]}, which originates an additional
charged layer contributing to the external electric field. Applying
the same parallel plate capacitor model as before, we get an electrostatic
energy difference that can be written as

\begin{equation}
V=\left(2-\frac{n}{n_{\textrm{a}}}\right)\frac{e^{2}n_{\textrm{a}}d}{2\varepsilon_{r}\varepsilon_{0}}.\label{eq:VnUnSohta}\end{equation}
Following a similar reasoning to the case of exfoliated graphene on
top of SiO$_{2}$, we would write $\varepsilon_{r}\approx(\varepsilon_{\textrm{SiC}}+1)/2\approx5$.
However, this value neglects that interface states (the effective
bottom plate capacitor) occur above the SiC substrate, close to the
graphene system, and thus $\varepsilon_{r}\approx1$ should be more
appropriate. In Fig.~\ref{fig:device}(c) we compare Eq.~\eqref{eq:VnUnSohta}
with experimental results for $V$ obtained by fitting ARPES measurements
from Ref.~\onlinecite{OBS+06}. For this particular biased BLG realization,
the as-prepared carrier density was $n_{\textrm{a}}\approx10^{13}\,\mbox{cm}^{-2}$.
From Eq.~\eqref{eq:VnUnSohta}, this $n_{\mbox{a}}$ value implies
a zero $V$, i.e., zero electric field and therefore zero gap, for
the bilayer density $n^{\textrm{th}}\approx2\times10^{13}\,\mbox{cm}^{-2}$.
Experimentally, a zero gap was found around $n^{\textrm{exp}}\approx2.3\times10^{13}\,\mbox{cm}^{-2}$.
Given the simplicity of the theory, it can be said that $n^{\textrm{th}}$
and $n^{\textrm{exp}}$ are in good agreement. However, the agreement
is only good at $V\sim0$, since the measured $V$ is not a linear
function of $n$, as Eq.~\eqref{eq:VnUnSohta} implies. In what follows
we analyze in detail the effect of screening and how it modifies Eqs.~\eqref{eq:VnUnSgeim}
and~\eqref{eq:VnUnSohta}.


\subsubsection{Screening correction}

\label{subsubsec:screen2}

In deriving Eqs.~\eqref{eq:VnUnSgeim} and~\eqref{eq:VnUnSohta}
we assumed that the electric field $E$ in the BLG region was exactly
the external one, $E_{ext}$. There is, however, an obvious additional
contribution: the external electric field polarizes the BLG, inducing
some charge asymmetry between the two graphene layers, which in turn
give rise to an internal electric field, $E_{int}$, that screens
the external one.

To estimate $E_{int}$ we can again apply a parallel plate capacitor
model. The internal electric field due to the charge asymmetry between
planes may thus be written as\begin{equation}
E_{int}=\frac{e\Delta n}{2\varepsilon_{r}\varepsilon_{0}},\label{eq:Eint}\end{equation}
where $-e\Delta n$ is the induced charge imbalance between layers,
which can be estimated through the weight of the wave functions in
each layer,\begin{multline}
\Delta n=n_{1}-n_{2}=\frac{2}{N_{\textrm{c}}A_{\hexagon}}\sum_{j,l=\pm}\sideset{}{'}\sum_{\mathbf{k}}\\
\big(|\varphi_{A1,\mathbf{k}}^{jl}|^{2}+|\varphi_{B1,\mathbf{k}}^{jl}|^{2}-|\varphi_{A2,\mathbf{k}}^{jl}|^{2}-|\varphi_{B2,\mathbf{k}}^{jl}|^{2}\big),\label{eq:Dndef}\end{multline}
where the factor 2 comes from spin degeneracy, $N_{\textrm{c}}$ is
the number of unit cells and $A_{\hexagon}=a^{2}\sqrt{3}/2$ is the
unit cell area, $jl$ is a band label, and the prime sum runs over
all occupied $\mathbf{k}$'s in the first BZ. The amplitudes $\varphi_{Ai,\mathbf{k}}^{jl}$
and $\varphi_{Bi,\mathbf{k}}^{jl}$, with $i=1,2$, are determined
by diagonalization of Eq.~\eqref{eq:HkBilayer}, enabling $\Delta n$
to be written as\begin{multline}
\Delta n=\frac{2}{N_{\textrm{c}}A_{\hexagon}}\sum_{j,l=\pm}\sideset{}{'}\sum_{\mathbf{k}}\\
\frac{(\epsilon_{\mathbf{k}}^{2}+\mathcal{K}_{\mathbf{k},-}^{jl})(\epsilon_{\mathbf{k}}^{2}-\mathcal{K}_{\mathbf{k},+}^{jl})^{2}-(\epsilon_{\mathbf{k}}^{2}+\mathcal{K}_{\mathbf{k},+}^{jl})t_{\perp}^{2}\mathcal{K}_{\mathbf{k},-}^{jl}}{(\epsilon_{\mathbf{k}}^{2}+\mathcal{K}_{\mathbf{k},-}^{jl})(\epsilon_{\mathbf{k}}^{2}-\mathcal{K}_{\mathbf{k},+}^{jl})^{2}+(\epsilon_{\mathbf{k}}^{2}+\mathcal{K}_{\mathbf{k},+}^{jl})t_{\perp}^{2}\mathcal{K}_{\mathbf{k},-}^{jl}},\label{eq:Dnk}\end{multline}
where $\epsilon_{\mathbf{k}}$ is the SLG dispersion, $\mathcal{K}_{\mathbf{k},\pm}^{jl}=(V/2\pm E_{\mathbf{k}}^{jl})^{2}$
with $E_{\mathbf{k}}^{jl}$ given by Eq.~\eqref{eq:Ekbias}. Taking
the limit $N_{\textrm{c}}\rightarrow\infty$, it is possible to write
Eq.~\eqref{eq:Dnk} as an energy integral weighted by the density
of states of SLG, as described in Appendix~\ref{secap:Dn}. What
is important to note is that in order to calculate $\Delta n$ we
must specify $V$, which in turn depend $\Delta n$ through Eq.~\eqref{eq:Eint}.
Thus, a self-consistent procedure must be followed. In particular,
for the two experimental realizations of biased BLG discussed in Sec.~\ref{subsubsec:screen1},
the self-consistent equation that determines $V$ reads: in the case
of exfoliated BLG,\citep{CNM+06}\begin{equation}
V=\left[\frac{n}{n_{0}}-2+\frac{\Delta n(n,V)}{n_{0}}\right]\frac{e^{2}n_{0}d}{2\varepsilon_{r}\varepsilon_{0}};\label{eq:VnSgeim}\end{equation}
in the case of epitaxial BLG,\citep{OBS+06} \begin{equation}
V=\left[2-\frac{n}{n_{\textrm{a}}}+\frac{\Delta n(n,V)}{n_{\textrm{a}}}\right]\frac{e^{2}n_{\textrm{a}}d}{2\varepsilon_{r}\varepsilon_{0}}.\label{eq:VnSohta}\end{equation}

\begin{figure}
\begin{centering}
\includegraphics[width=0.98\columnwidth]{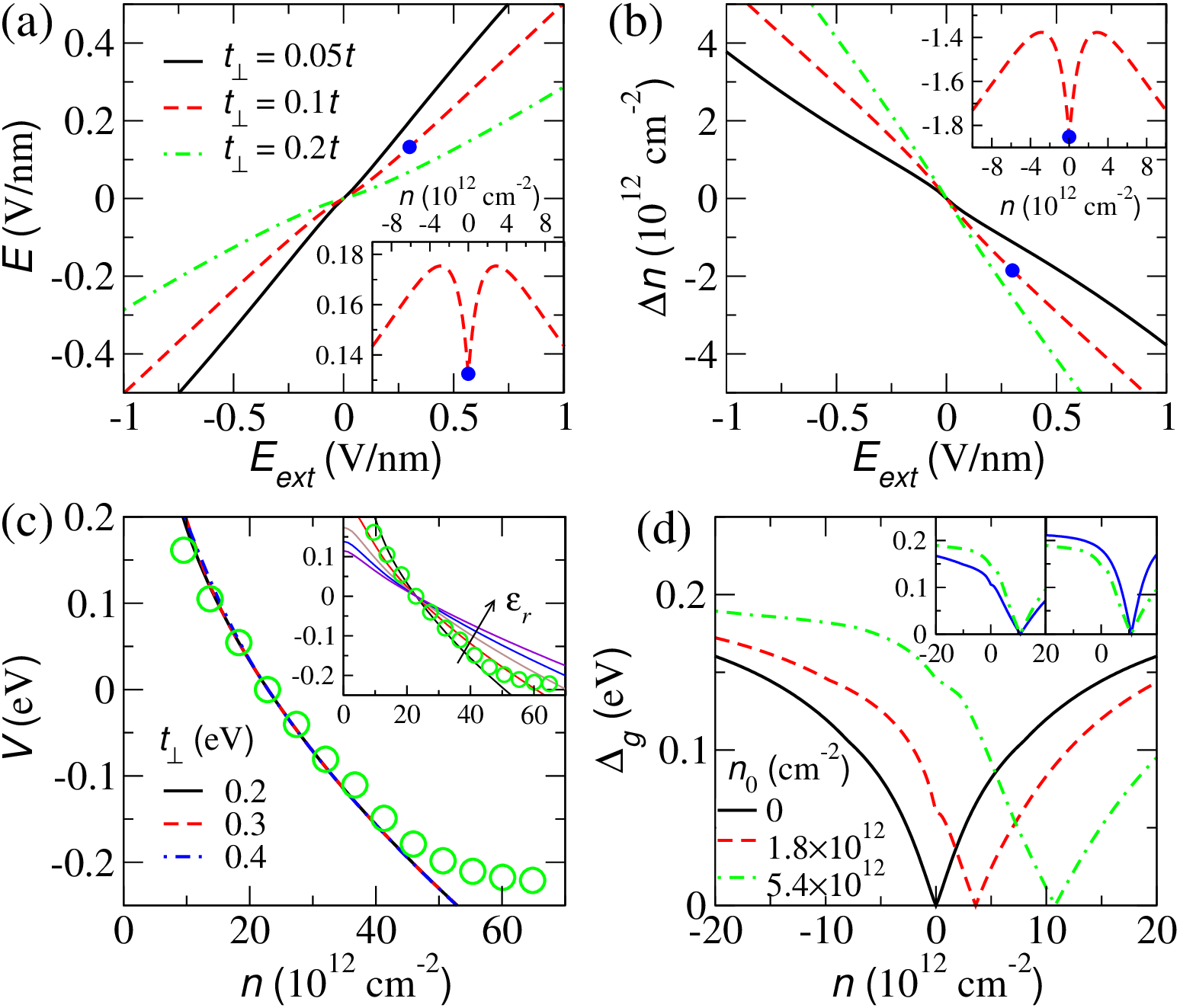}
\par\end{centering}

\caption{\label{fig:screening}(Color online) (a)-(b)~Respectively, screened
electric field and charge imbalance vs $E_{ext}$ at half-filling;
the insets show the effect of changing $n$ at fixed $E_{ext}=0.3\,\mbox{V/nm}$,
signaled by the (blue) dot in main panels. (c)~$V$ vs $n$ for the
BLG device shown in the right panel of Fig.~\ref{fig:device}(a):
symbols are experimental data from Ref.~\onlinecite{OBS+06}; lines
are the result of Eq.~\eqref{eq:VnSohta} for $\varepsilon_{r}=1$;
the effect of changing $\varepsilon_{r}=1-5$ is shown in the inset.
(d)~Gap vs $n$ for the BLG device shown in the left panel of Fig.~\ref{fig:device}(a)
with $t_{\perp}\simeq0.22\,\mbox{eV}$ and $\varepsilon_{r}=1$; the
left inset compares the $n_{0}=5.4\times10^{12}\,\mbox{cm}^{-2}$
result for $\varepsilon_{r}=1$ (green dashed-dotted) with $\varepsilon_{r}=2$
(blue full line); the right inset shows the $n_{0}=5.4\times10^{12}\,\mbox{cm}^{-2}$
result for the screened $V$ given by Eq.~\eqref{eq:VnSgeim} (dashed-dotted
line) and for the unscreened $V$ given by Eq.~\eqref{eq:VnUnSgeim}
(full line). We used as in-plane hopping $t\simeq3\,\mbox{eV}$.}

\end{figure}

The self-consistent electric field $E=E_{ext}+E_{int}$ at the BLG
region, with $E_{int}$ given by Eq.~\eqref{eq:Eint} for $\varepsilon_{r}=1$,
is shown at half-filling as a function of $E_{ext}$ in Fig.~\ref{fig:screening}(a).
The screened $E$ is approximately a linear function of $E_{ext}$,
with a constant of proportionality that depends on the specific value
of $t_{\perp}$. Increasing $t_{\perp}$ leads to an increased screening,
which can be understood as due to an increased charge imbalance between
layers, as shown in Fig.~\ref{fig:screening}(b). The highly non-linear
effect of inducing a finite carrier density ($n\neq0$) can be seen
in the insets of Fig.~\ref{fig:screening}(a) and~\ref{fig:screening}(b),
for $t_{\perp}=0.1t$ and $E_{ext}=0.3\,\mbox{V/nm}$.

As a validation test to the present self-consistent treatment, we
compare Eq.~\eqref{eq:VnSohta} with experimental results for $V$
obtained by fitting ARPES measurements from Ref.~\onlinecite{OBS+06},
as mentioned in Sec.~\ref{subsubsec:screen1}. The result is shown
in Fig.~\ref{fig:screening}(c). Clearly, the self-consistent $V$
given by Eq.~\eqref{eq:VnSohta} for $\varepsilon_{r}=1$ is a much
better approximation than the unscreened result of Eq.~\eqref{eq:VnUnSohta}
{[}see Fig.~\ref{fig:device}(c){]}. The best fit is obtained for
$\varepsilon_{r}\sim1-2$, as can be seen in the inset of Fig.~\ref{fig:screening}(c).
The value $\varepsilon_{r}\approx(\varepsilon_{\textrm{SiC}}+1)/2\approx5$
is too high, possibly because the bottom capacitor plate is, indeed,
due to interface states,\citep{foot3} and therefore is not buried
inside the SiC substrate.\citep{VFH+07,MP07,MVN+07} Note, however,
that the dielectric constant $\varepsilon_{r}$ may effectively be
tuned externally, as recently shown in SLG by adding a water overlayer
in ultra-high vacuum.\citep{JAC+08} In Fig.~\ref{fig:screening}(d)
we show the gap $\Delta_{g}$ as a function of carrier density $n$
for the biased BLG device shown in the left panel of Fig.~\ref{fig:device}(a),
with realistic values of chemical doping $n_{0}$.\citep{CNM+06}
The gap is given by Eq.~\eqref{eq:gapV}, with $t_{\perp}\simeq0.22\,\mbox{eV}$\citep{CNM+06}
and $V$ obtained by solving self-consistently Eq.~\eqref{eq:VnSgeim}
for $\varepsilon_{r}=1$. Note that for $E_{ext}=0$ we always have
$E_{int}=0$ (the charge imbalance must be externally induced), and
therefore we also have $V=0$ and $\Delta_{g}=0$. For this particular
biased BLG device the present model predicts $E_{ext}=0$ for $n=2n_{0}$,
which explains the asymmetry for $\Delta_{g}$ vs $n$ shown in Fig.~\ref{fig:screening}(d).

The most important characteristic of such devices, from the point
of view of applications, is the maximum size of the gap which could
be induced. The maximum $\Delta_{g}$ occurs when $V_{g}$ reaches
its maximum, which occurs just before the breakdown of $\mbox{SiO}_{2}$.
The breakdown field for $\mbox{SiO}_{2}$ is $\gtrsim1\mbox{V/nm}$,
meaning that $V_{g}$ values as high as $300\,\mbox{V}$ are possible
for the device shown in the left panel of Fig.~\ref{fig:device}(a).
From Eq.~(\ref{eq:nVg}) we see that $V_{g}\simeq\pm300\,\mbox{V}$
implies $n-n_{0}\simeq\pm22\times10^{12}\,\mbox{cm}^{-2}$, and therefore
Fig.~\ref{fig:screening}(d) nearly spans the interval of possible
densities. It is apparent, specially for $n_{0}=5.4\times10^{12}\,\mbox{cm}^{-2}$,
that when the maximum allowed densities are reached the gap seems
to be approaching a saturation limit. This saturation is easily identified
with the plateau shown in Fig.~\ref{cap:bandstructV}(b) for $\Delta_{g}$
vs $V$, occurring for $V\gtrsim t_{\perp}$. We may then conclude
that such devices enable the entire range of allowed gaps (up to $t_{\perp}$)
to be accessed --- as has been shown in very recent experiments.\citealp{MLS+09,ZTG+09}
The effect of using a different dielectric constant ($\varepsilon_{r}=2$)
is shown as a full line in the left inset of Fig.~\ref{fig:screening}(d),
and the result for the unscreened case in the right inset, both for
$n_{0}=5.4\times10^{12}\,\mbox{cm}^{-2}$. The former makes the gap
slightly smaller, and the latter slightly larger, but the main conclusions
remain.

\subsubsection{Screening in continuum models}

\label{subsubsec:GrapheneBEPscreenCont}

The self consistent Hartree approach considered in the previous section
has been applied to the full tight-binding Hamiltonian given in Eq.~\eqref{eq:Hbilayer}.
Here we compare the results for the potential difference $V$ and
gap $\Delta_{g}$ when the screening correction is used within the
continuum approximation, either for the 4-band model of Eq.~\eqref{eq:HKbl}
or for the 2-band model of Eq.~\eqref{eq:Heff}. This self consistent
Hartree approach in the continuum has been followed in Refs.~\onlinecite{NNG+06,McC06}.

In the case of the 4-band model, $\Delta n$ is still given by Eq.~\eqref{eq:Dnk}
with the substitutions $\epsilon_{\mathbf{k}}\to v_{\textrm{F}}p$
and $\frac{2}{N_{\textrm{c}}A_{\hexagon}}\sum_{j,l=\pm}\sideset{}{'}\sum_{\mathbf{k}}\to\frac{2}{\pi\hbar^{2}}\sideset{}{'}\sum_{j,l=\pm}\int_{p_{1}}^{p_{2}}\mbox{d}p\, p$,
where the prime on the right hand summation means sum over total or
partially occupied bands. Depending on the band in question and the
value of the Fermi energy $E_{\textrm{F}}$, the limits of integration
are $p_{1},p_{2}=\{0,p^{\pm},\Lambda\}$, where\begin{equation}
v_{\textrm{F}}p^{\pm}=\sqrt{E_{\textrm{F}}^{2}+V^{2}/4\pm\sqrt{E_{\textrm{F}}^{2}(V^{2}+t_{\perp}^{2})-t_{\perp}^{2}V^{2}/4}},\label{eq:kpmLim}\end{equation}
and $\Lambda$ is a BZ cutoff that can be chosen such that $\frac{4\pi}{\hbar^{2}}\int_{0}^{\Lambda}\mbox{d}p\, p=\frac{4\pi^{2}}{A_{\hexagon}}\Leftrightarrow\Lambda=\hbar\sqrt{\pi/A_{\hexagon}}$.
As regards the gap $\Delta_{g}$, in the 4-band model it is still
given by Eq.~\eqref{eq:gapV}.

For the 2-band model case, the charge imbalance can be written as
an integral in momentum space of the function $|\phi_{B1}|^{2}-|\phi_{A2}|^{2}=\pm V/(V^{2}+4v_{\textrm{F}}^{4}p^{4}/t_{\perp}^{2})^{1/2}$,
where $\Phi=(\phi_{B1},\phi_{A2})$ is the two component wave function
obtained by diagonalizing Eq.~\eqref{eq:Heff}. The $\pm$ signs
stand for the contribution of valence and conduction bands, respectively.
In particular, at half-filling the charge imbalance is given by \begin{equation}
\Delta n_{1/2}\simeq-\frac{t_{\perp}V}{2\pi v_{\textrm{F}}^{2}\hbar^{2}}\ln\bigl(2t_{\perp}/|V|+\sqrt{4t_{\perp}^{2}/V^{2}+1}\bigr),\label{eq:Dn2bHalf}\end{equation}
where we have included a factor of 4 to account for both spin and
valley degeneracies. The BZ cutoff $\Lambda$ has been chosen such
that $v_{\textrm{F}}\Lambda=t_{\perp}$.\citep{McC06} Since in the
2-band model it is assumed that $V\ll t_{\perp}$ holds we can write
$\Delta_{1/2}\approx-t_{\perp}V/(2\pi v_{\textrm{F}}^{2}\hbar^{2})\ln(4t_{\perp}/|V|)$,
which, from Eq.~\eqref{eq:Eint}, leads to the logarithmic divergence
of the screening ratio at small external electric field, $E_{ext}/E\sim-\ln E$,
as mentioned in Ref.~\onlinecite{MSB+06}. For a general filling
$n$ the charge imbalance reads\begin{equation}
\Delta n\approx\frac{t_{\perp}V}{2\pi v_{\textrm{F}}^{2}\hbar^{2}}\ln\biggl(\frac{v_{\textrm{F}}^{2}\hbar^{2}\pi|n|}{2t_{\perp}^{2}}+\sqrt{\frac{v_{\textrm{F}}^{4}\hbar^{4}\pi^{2}n^{2}}{4t_{\perp}^{4}}+1}\biggr),\label{eq:Dn2b}\end{equation}
where the charge density is given in terms of the Fermi wave vector
as $n=\pm p_{\textrm{F}}^{2}/(\pi\hbar^{2})$. Inserting Eq.~\eqref{eq:Dn2b}
into Eq.~\eqref{eq:VnSgeim} or~\eqref{eq:VnSohta} we get the expression
for $V$ in the 2-band approximation, which is exactly the gap in
the 2-band model, $\Delta_{g}=|V|$.

\begin{figure}
\begin{centering}
\includegraphics[clip,width=0.98\columnwidth]{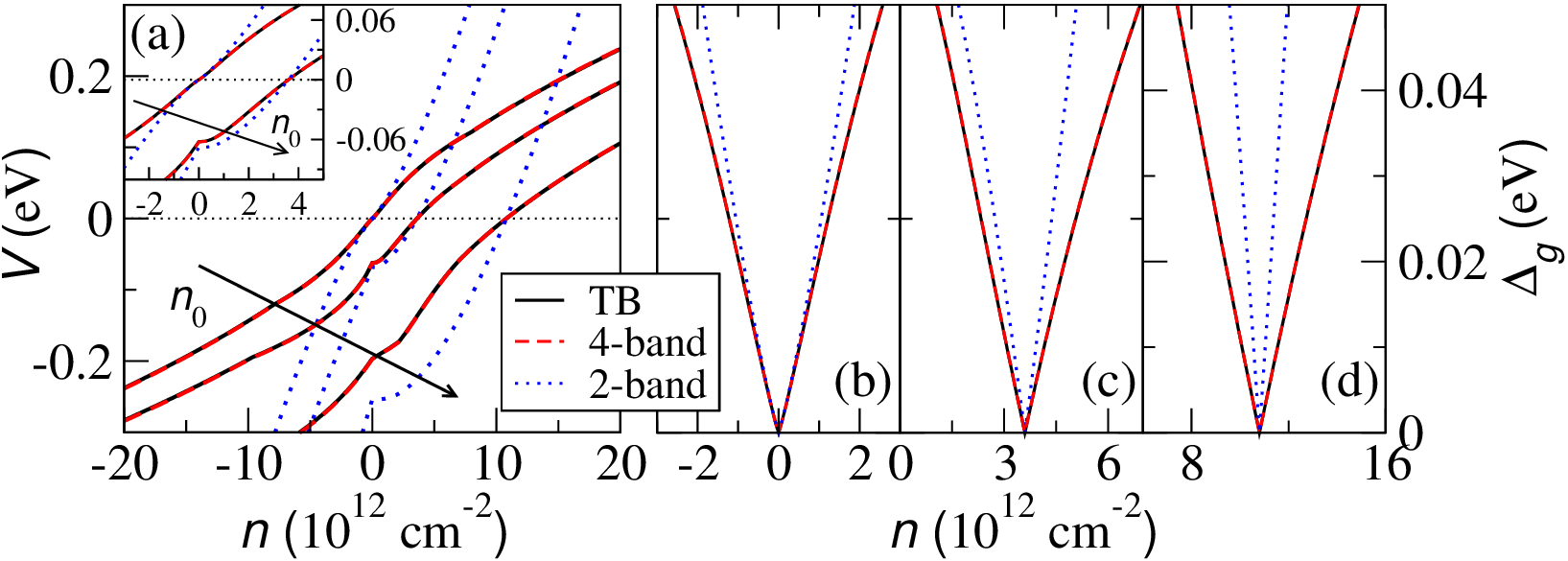}
\par\end{centering}

\caption{\label{fig:comparLattCont}(Color online) (a)~Screened $V$ vs $n$
for the BLG system shown in the left panel of Fig.~\ref{fig:device}(a)
computed within three different approaches (see text): full tight-binding
(TB), 4-band approximation, and 2-band approximation. Three different
chemical dopings have been considered, $n_{0}=\{0,1.8,5.4\}\times10^{12}\,\mbox{cm}^{-2}$.
The inset shows a zoom around $V=0$ for $n_{0}=\{0,1.8\}\times10^{12}\,\mbox{cm}^{-2}$.
(b)-(c)~Screened gap vs $n$ obtained using $V$ shown in~(a), respectively
for $n_{0}=\{0,1.8,5.4\}\times10^{12}\,\mbox{cm}^{-2}$. Parameters:
$t\simeq3\,\mbox{eV}$, $t_{\perp}\simeq0.22\,\mbox{eV}$, and $\varepsilon_{r}=1$.}

\end{figure}

In Fig.~\ref{fig:comparLattCont}(a) the obtained electrostatic energy
difference between planes $V$ is shown for the three different approaches
discussed above. The full (black) lines stand for the full tight-binding
result, with $V$ given by Eq.~\eqref{eq:VnSgeim} and the charge
imbalance $\Delta n$ by Eq.~\eqref{eq:Dnk}. The result obtained
in the 4-band approximation is shown as dashed (red) lines. It can
hardly be distinguished from the full tight-binding result, even when
the chemical doping $n_{0}$ is as high as $5.4\times10^{12}\,\mbox{cm}^{-2}$
(see figure caption). In fact, the only prerequisite for the continuum
4-band approximation {[}Eq.~\eqref{eq:HKbl}{]} to hold is that $|E_{\textrm{F}}|\ll t$,
which is always realized for the available BLG devices. As regards
the 2-band approximation model, we show as dotted (blue) lines the
self-consistent result for $V$, obtained fro $\Delta n$ as in Eq.~\eqref{eq:Dn2b}
. Clearly, it is only when both the bilayer density $n$ and the chemical
doping $n_{0}$ are small enough for the relation $|E_{\textrm{F}}|,V\ll t_{\perp}$
to hold that the 2-band model is a good approximation (see inset).
The same conclusions apply to the behavior of the gap $\Delta_{g}$
as a function of carrier density $n$, which is shown in panels \ref{fig:comparLattCont}(b)-(d)
for $n_{0}=\{0,1.8,5.4\}\times10^{12}\,\mbox{cm}^{-2}$, respectively.
The failure of the 2-band model in the presence of interactions was
also observed in Hartree calculations of the electron compressibility.\citep{KNC+08}

\subsubsection{Electron-hole asymmetry}

\label{subsubsec:GrapheneBEPscreenEHasym}

As we have seen in Secs.~\ref{subsubsec:screen1} and~\ref{subsubsec:screen2},
the two biased BLG devices shown in Fig.~\ref{fig:device}(a) have
zero gap when the carrier density is twice the system's chemical doping.
The closing of the gap at a finite density induces an electron-hole
asymmetric behavior in the system, where obvious examples are the
gap $\Delta_{g}$ and the electrostatic energy difference between
layers $V$, as shown in Figs.~\ref{fig:device}(d) and~\ref{fig:comparLattCont}(a).
An experimental confirmation for this electron-hole asymmetric behavior
comes from measurements of the cyclotron mass in the biased BLG device
shown in the left panel of Fig.~\ref{fig:device}(a)\citep{CNM+06}
(discussed in more detail in Sec.~\ref{subsec:BilayerMFEmc}). However,
real electron-hole asymmetry can also be present in BLG due to extra
hopping terms, as mentioned in Sec.~\ref{sec:BilayerModel}. Here
we study how $\Delta_{g}$ and $V$ are effected by the electron-hole
symmetry breaking terms $t'$, $\gamma_{4}$, and $\Delta$, taking
into account the screening correction.

\begin{figure}
\noindent \begin{centering}
\includegraphics[width=0.8\columnwidth]{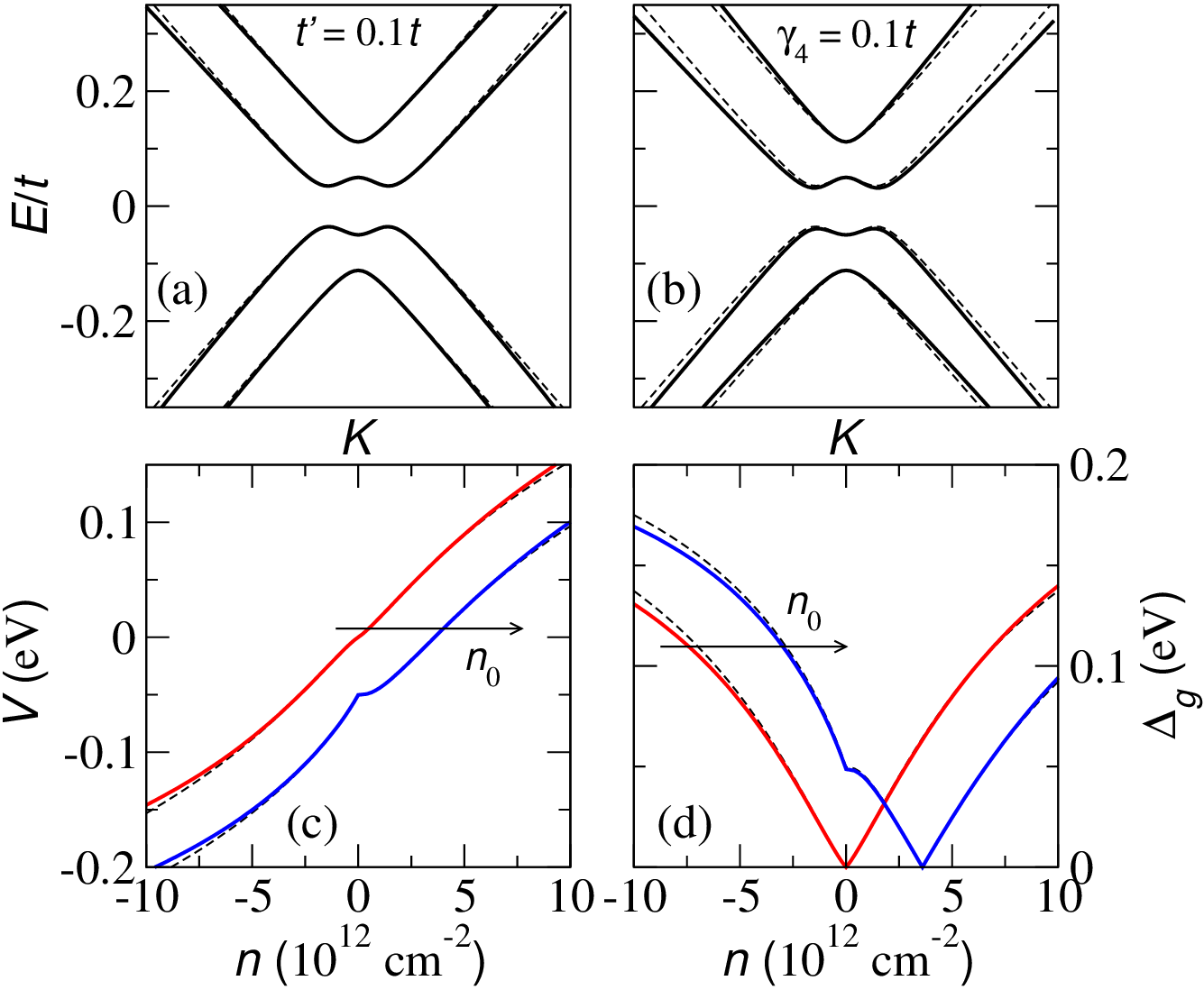}
\par\end{centering}

\caption{\label{fig:ehAsym}(Color online) (a)-(b)~Band structure around $K$
for the biased BLG with $t'=0.1t$ and $\gamma_{4}=0.1t$, respectively,
for $V=t_{\perp}=0.1t$. Dashed lines: $t'=\gamma_{4}=0$. (c)-(d)~Respectively,
$V$ vs $n$ and $\Delta_{g}$ vs $n$ for the BLG device shown in
the left panel of Fig.~\ref{fig:device}(a), modeled with a finite
$\gamma_{4}$. Parameters: $t\simeq3\,\mbox{eV}$, $t_{\perp}=0.1t$,
$\gamma_{4}=0.1t$, $\varepsilon_{r}=1$, and $n_{0}=\{0,1.8\}\times10^{12}\,\mbox{cm}^{-2}$.
Dashed lines: $t'=\gamma_{4}=0$.}

\end{figure}

Inclusion of in-plane second-NN hopping $t'$ leads to a generalized
version of Eq.~\eqref{eq:HkBilayer}, which can be written as $H_{\mathbf{k},t'}=H_{\mathbf{k}}-(\epsilon_{\mathbf{k}}^{2}t'/t-3t')\mathbf{1}$,
where $H_{\mathbf{k}}$ is given by Eq.~\eqref{eq:HkBilayer}, $\epsilon_{\mathbf{k}}$
is the SLG dispersion, and $\mathbf{1}$ is the $4\times4$ identity
matrix. The generalized the BLG dispersion, either biased or unbiased,
is given by the $t'=0$ result added by $-\epsilon_{\mathbf{k}}^{2}t'/t+3t'$,
which clearly breaks electron-hole symmetry. Note that a finite $t'$
has no influence on the wavefunctions' amplitude. Therefore, the integrand
in Eq.~\eqref{eq:Dndef} -- the definition of the charge carrier
imbalance between layers $\Delta n$ -- is independent of $t'$. We
have found numerically, using a 4-band continuum model, that neither
the screened $V$ nor the gap $\Delta_{g}$ are affected by $t'$,
although the gap becomes indirect for finite $t'$. This means that
the structure of occupied $\mathbf{k}$'s is insensitive to $t'$,
and thus $\Delta n$ in Eq.~\eqref{eq:Dndef} is fully $t'$ independent,
at least as long as $E_{\textrm{F}}\ll t$. Even though the presence
of $t'$ can lead to the suppression of the Mexican hat in the valence
band, this only happens for $|V|<t_{\perp}^{2}t'\sim10^{-3}t$. For
such a small $|V|$ value the Mexican hat plays an irrelevant role.
The band structure around the $K$ point for $t'=0.1t$ (solid line)
and $t'=0$ (dashed line) can be seen in Fig.~\ref{fig:ehAsym}(a)
for typical parameter values.

Now we turn to the effect of the inter-layer second-NN hopping $\gamma_{4}$.
The generalized version of Eq.~\eqref{eq:HkBilayer} for finite $\gamma_{4}$,
which we call $H_{\mathbf{k},\gamma_{4}}$, can be obtained by replacing
the null entries $(A1,A2)$ and $(B1,B2)$ by $\gamma_{4}s_{\mathbf{k}}^{*}$
and $(A2,A1)$ and $(B2,B1)$ by $\gamma_{4}s_{\mathbf{k}}$. The
associated eigenproblem admits an analytic treatment at low energies
and small biases $v_{\textrm{F}}p,V\ll t_{\perp}$,\citep{NNP+05}
but as has been seen previously $V\sim t_{\perp}$ is possible in
real systems. Therefore, we analyze the problem numerically using
a 4-band continuum approximation. The matrix Hamiltonian $H_{\mathbf{k},\gamma_{4}}$
may then be written as $H_{K,\gamma_{4}}=M^{\dagger}\tilde{H}_{K,\gamma_{4}}M$
near the $K$ points, with $M=\mbox{diag}[1,e^{i\varphi_{\mathbf{p}}},e^{-i\varphi_{\mathbf{p}}},1]$,
and $\tilde{H}_{K,\gamma_{4}}$ obtained from Eq.~\eqref{eq:HKbl}
with $\varphi_{\mathbf{p}}=0$ and the null entries $(A1,A2)$, $(B1,B2)$,
$(A2,A1)$, and $(B2,B1)$ replaced by $-v_{4}p$, where $v_{4}=\gamma_{4}a\hbar^{-1}\sqrt{3}/2\lesssim10^{5}\,\mbox{ms}^{-1}$.
The canonical transformation defined by $M$ clearly shows that the
problem still has cylindrical symmetry in the continuum approximation.
Around the $K'$ points we have $H_{K',\gamma_{4}}=M\tilde{H}_{K,\gamma_{4}}M^{\dagger}$.
The obtained band structure for $\gamma_{4}=0.1t$ (solid lines) and
$\gamma_{4}=0$ (dashed lines) is shown in Fig.~\ref{fig:ehAsym}(b)
for typical parameter values. Note that, even though the gap becomes
indirect for $\gamma_{4}\neq0$, we still have $E_{p=0}=\{\pm V/2,\pm\sqrt{t_{\perp}^{2}+V^{2}/4}\}$
as in the $\gamma_{4}=0$ case. The screened electrostatic energy
difference between layers $V$ for the biased BLG device shown in
the left panel of Fig.~\ref{fig:device}(a) is shown as a function
of the carrier density in Fig.~\ref{fig:ehAsym}(c). The result for
$V$ has been obtained by solving Eq.~\eqref{eq:VnSgeim} with carrier
imbalance $\Delta n$ given by the continuum version of Eq.~\eqref{eq:Dndef},
with wavefunctions obtained numerically through $\tilde{H}_{K,\gamma_{4}}$
for $\gamma_{4}=0.1t$ (see figure caption for other parameter values).
The corresponding screened gap $\Delta_{g}$ is shown in panel~\ref{fig:ehAsym}(d).
The $\gamma_{4}=0$ result is also shown as a dashed line for both
$V$ and $\Delta_{g}$. The effect of $\gamma_{4}$ may clearly be
considered small, even for such a large value as $\gamma_{4}\simeq0.3\,\mbox{eV}$.
However, electronic properties which are particularly sensitive to
the changes of the Fermi surface (like, for instance, the cyclotron
mass), may, in principle, be measurably affected by $\gamma_{4}$.
We will come back to this point in Sec.~\ref{subsec:BilayerMFEmc}.

As regards the on-site energy $\Delta$, since it is smaller than
$\gamma_{4}$ (see Sec.~\ref{sec:BilayerModel}) we consider their
simultaneous effect. The additional term in the Hamiltonian adds to
the matrix $H_{\mathbf{k},\gamma_{4}}$ the contribution $\mbox{diag}[\Delta,0,0,\Delta]$,
and therefore the 4-band continuum approximation for finite $\gamma_{4}$
and $\Delta$ may be written as $\tilde{H}_{K,\gamma_{4},\Delta}=\tilde{H}_{K,\gamma_{4}}+\mbox{diag}[\Delta,0,0,\Delta]$,
where we use the same transformation $M$ introduced above. Similarly
to $\gamma_{4}$, the effect of $\Delta$ is negligible in both $V$
and $\Delta_{g}$.

\subsection{DOS and LDOS}

\label{subsec:BilayerBEPdos}

Insight into the electronic properties of biased (and unbiased) BLG
can also be achieved by studying the density of states (DOS) and the
local DOS (LDOS) of the system. In particular, the LDOS can be accessed
through scanning tunneling microscopy/spectroscopy measurements,\citep{TH85}
providing a useful way to validate theoretical models. On the other
hand, the knowledge of the DOS turns out to be very useful for practical
purposes, as it provides a way to relate the Fermi energy $E_{\textrm{F}}$
and the carrier density $n$ in the system: $|n|=\int_{0}^{|E_{\textrm{F}}|}\textrm{d}E\,\rho_{2}(E)$,
where $\rho_{2}(E)$ stands for the BLG DOS. 

We have computed the analytical expression for the DOS of BLG, valid
over the entire energy spectrum and for zero and finite bias. The
expression is given in Appendix~\ref{secap:BilayerDOS}. As regards
the LDOS, the results have been obtained using the recursive Green's
function method.\citep{Hayd80} The DOS and LDOS of unbiased BLG has
been obtained previously within the effective mass approximation in
Ref.~\onlinecite{WLS+06}. The effect of disorder on the DOS and
LDOS of BLG, both biased and unbiased, has also been studied recently.\citep{NNGP06,NN06,CPS06,WLS+06,Bena07,NNG+07}

\begin{figure}
\begin{centering}
\includegraphics[clip,width=0.9\columnwidth]{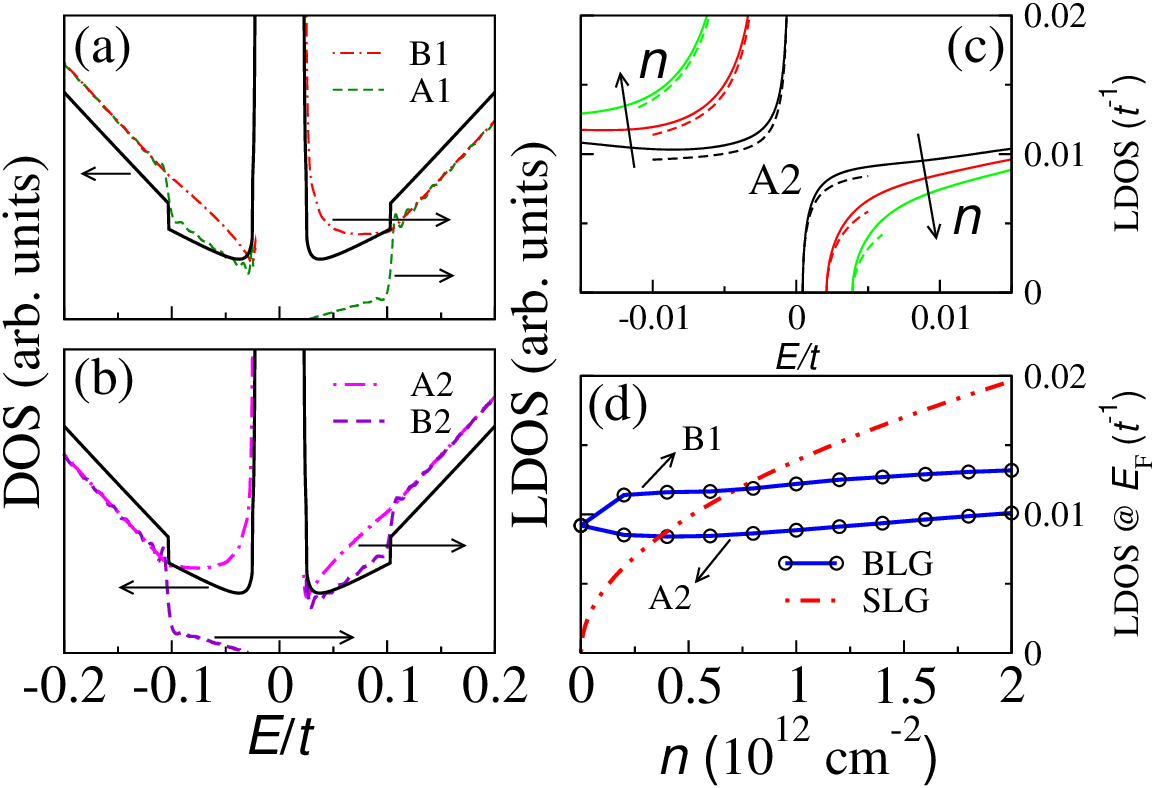}
\par\end{centering}

\caption{\label{cap:DOS}(Color online) (a)-(b)~LDOS of BLG at $A1/B1$ and
$A2/B2$ sites, respectively, for $V=0.05t$ and $t_{\perp}=0.1t$.
The total DOS is shown as a full line. (c)~LDOS at $A2$ sites for
$n\simeq\{0.2,0.8,1.4\}\times10^{12}\,\mbox{cm}^{-2}$ and $t_{\perp}=0.1t$.
Full lines for numerical results and dashed lines for Eq.~\eqref{eq:dos2bm}.
(d)~LDOS at $E_{\textrm{F}}$ vs $n$ for BLG and SLG.}

\end{figure}

The DOS (full line) and LDOS (dashed and dash-dotted lines) for the
biased BLG is shown in Fig.~\ref{cap:DOS}(a)-(b) for $V=0.05t$.
The asymmetry between the four sublattices is evident, in particular
between sites $B1$ and $A2$, and $A1$ and $B2$, which are equivalent
in the unbiased system. Note that close to the gap edges the states
corresponding to positive energies have a larger amplitude at $B1$
sites, while those corresponding to negative energies have a larger
amplitude at $A2$ sites. This behavior agrees with the observation
that $B1$ and $A2$ are the low energy active sites (the basis for
the 2-band model), and it also reflects our choice of electrostatic
energies in Eq.~\eqref{eq:HV}: $+V/2$ in layer~1 and $-V/2$ in
layer~2. The asymmetry between $B1$ and $A2$ sites can be understood
with the 2-band continuum model, valid for $v_{\textrm{F}}p,V\ll t_{\perp}$.
Defining the LDOS as $\rho_{B1/A2}(E)=\frac{1}{N_{\textrm{c}}}\sum_{\mathbf{k}}|\phi_{B1/A2,\mathbf{k}}|^{2}\delta(E-E_{\mathbf{k}})$,
where $\Phi_{\mathbf{k}}=(\phi_{B1,\mathbf{k}},\phi_{A2,\mathbf{k}})$
is the two component wave function obtained by diagonalizing Eq.~\eqref{eq:Heff},
we can readily arrive at the following expressions,\begin{equation}
\rho_{B1/A2}(E)=\frac{1}{2\sqrt{3}\pi}\frac{t_{\perp}}{t^{2}}\mbox{sgn}(E)\frac{E\pm V/2}{\sqrt{E^{2}-V^{2}/4}}.\label{eq:dos2bm}\end{equation}
The asymmetric behavior is apparent, with $\rho_{B1}(E)$ diverging
for $E\to V/2^{+}$ and $\rho_{A2}(E)$ for $E\to-V/2^{-}$. The result
for $\rho_{A2}(E)$ is shown in Fig.~\ref{cap:DOS}(c) for $V\simeq\{0.87,4.23,7.87\}\times10^{-3}t$
and $t_{\perp}=0.1t$. Within the screening corrected parallel plate
capacitor model discussed in Sec.~\ref{subsec:screen} {[}Eq.~\eqref{eq:VnSgeim}{]},
these $V$ values correspond to carrier densities $n\simeq\{0.2,0.8,1.4\}\times10^{12}\,\mbox{cm}^{-2}$,
respectively, where we have used $t\simeq3.1\,\mbox{eV}$, $n_{0}=0$,
and $\varepsilon_{r}=1$. The full lines are the recursive Green's
function method\citep{Hayd80} results and dashed lines are the results
of Eq.~\eqref{eq:dos2bm}. As expected, the closer to the gap edges
the better the agreement between the two approaches.

A strong suppression of electrical noise in BLG has been reported
recently by Lin and Avouris.\citep{LA08} In devices made from exfoliated
BLG on top of $\mbox{SiO}_{2}$, the current fluctuations are thought
to originate from the fluctuating trapped charges in the oxide. Therefore,
the more effective the impurity charge screening in the system the
lower the electrical noise. The lower noise in BLG than in SLG may
then be attributed to the low energy finite DOS in the former. However,
it has also been reported in Ref.~\onlinecite{LA08} that while increasing
the carrier density in SLG leads to lower noise, as expected due to
more effective impurity screening, it results in higher noise in BLG.
Insight into this behavior is achieved by analyzing the LDOS at the
Fermi level $E_{\textrm{F}}$ in a biased BLG, as charging the system
through the back gate $V_{g}$ leads to a finite perpendicular electric
field. In Fig.~\ref{cap:DOS}(e) we show the biased BLG LDOS at $E_{\textrm{F}}$
for $B1$ and $A2$ sites as a function of carrier density $n$ in
the system. For a given $n$, the electrostatic energy difference
$V$ is evaluated self-consistently through Eq.~\eqref{eq:VnSgeim},
with $n_{0}=0$ and $\varepsilon_{r}=1$, and $E_{\textrm{F}}$ is
obtained by integrating over the DOS. Additionally, we use $t\simeq3\,\mbox{eV}$
and $t_{\perp}=0.1t$. We have chosen densities in the range $n\in[0-2]\times10^{12}\,\mbox{cm}^{-2}$,
which corresponds to back gate voltages $V_{g}\in[0-27]\,\mbox{eV}$
through Eq.~\eqref{eq:nVg}, similar to the experimental range in
Ref.~\onlinecite{LA08}. The main observation to be made as regards
the results of Fig.~\ref{cap:DOS}(e) is that for the low energy
active sublattices $B1$ and $A2$ the LDOS at $E_{\textrm{F}}$ remains
approximately constant with increasing electron density, as opposed
to the $\sim\sqrt{n}$ dependence found in SLG. This is an indication
that impurity screening may not be increasing with carrier density
in the biased BLG system, which may be contributing to enhance electrical
noise.

\section{Magnetic field effects}

\label{sec:BilayerMFE}

In the biased BLG system, as a consequence of the gapped band structure
discussed in Sec.~\ref{sec:bulkelecprop}, a perpendicular magnetic
field is expected to induce distinct features in electronic properties.
In this section we focus on the cyclotron mass (semi-classical approach)
and on the cyclotron resonance (quantum regime) comparing the theory
with experimental results.

%

\subsection{Cyclotron mass}

\label{subsec:BilayerMFEmc}

In the semi-classical approximation the cyclotron effective mass $m_{\textrm{c}}$
is given by \begin{equation}
m_{\textrm{c}}=\frac{\hbar^{2}}{2\pi}\frac{\partial A(E)}{\partial E}\Big|_{E=E_{\textrm{F}}},\label{eq:mcdef}\end{equation}
where $A(E)$ is the $k$-space area enclosed by the orbit of energy
$E$, and the derivative is evaluated at the Fermi energy $E_{\textrm{F}}$.\citep{LKinEFS80,LL9,ZimanPTS}
It can be accessed experimentally through the Shubnikov-de Haas effect,
providing a direct probe to the Fermi surface. In the case of exfoliated
graphene, either SLG or (un)biased BLG, the Fermi energy can be varied
by tuning the back gate voltage, and therefore a significant portion
of the whole band structure may be unveiled. In particular for the
biased BLG, the presence of a finite gap can be checked and the model
developed in Sec.~\ref{sec:bulkelecprop} tested.

\subsubsection{Comparison with experiment}

\label{subsec:BilayerMFEmcCompExp}

General expressions for $m_{\textrm{c}}$ obtained for the full tight-binding
bands in Eq.~\eqref{eq:Ekbias}, valid for the relevant parameter
range $V\lesssim t_{\perp}\ll t$ and restricted to $E_{\textrm{F}}<t$,
are given in Appendix~\ref{secap:Bilayermc}. In Fig.~\ref{cap:mc}(a)
we compare the theory results for the cyclotron mass with experimental
measurements\citep{CNM+06} on the biased BLG system shown in the
left panel of Fig.~\ref{fig:device}(a). We have only considered
$m_{\textrm{c}}$ associated with low energy bands $E_{\mathbf{k}}^{\pm-}$
{[}see Eq.~\eqref{eq:Ekbias}{]}, since $E_{\mathbf{k}}^{\pm+}$
are inactive for the experimentally available carrier densities. The
dashed lines stand for the unscreened result, where $V$ is given
by Eq.~\eqref{eq:VnUnSgeim}, and the solid lines are the screened
result, with $V$ given by Eq.~\eqref{eq:VnSgeim}. The inter-layer
coupling $t_{\perp}$ has been taken as an adjustable parameter, keeping
all other fixed: $t\simeq3\,\mbox{eV}$, $\varepsilon_{r}=1$, and
$n_{0}=1.8\times10^{12}\,\mbox{cm}^{-2}$. The value of $t_{\perp}$
could then be chosen so that theory and experiment gave the same $m_{\textrm{c}}$
for $n=2n_{0}\approx3.6\times10^{12}\textrm{cm}^{-2}$. As discussed
in Sec.~\ref{subsubsec:screen2}, at this particular density the
gap closes, meaning that the theoretical value becomes independent
of the screening assumptions. We found $t_{\perp}\approx0.22$~eV,
in good agreement with values found in the literature. The theoretical
dependence $m_{\textrm{c}}(n)$ agrees well with the experimental
data for the case of electron doping. Also, as seen in Fig.~\ref{cap:mc}(a),
the screened result provides a somewhat better fit than the unscreened
model, especially at low electron densities. This fact, along with
the good agreement found for the electrostatic energy difference data
of Ref.~\onlinecite{OBS+06} {[}see Fig.~\ref{fig:screening}(c){]},
allows us to conclude that for doping of the same sign from both sides
of bilayer graphene, the gap is well described by the screened approach.
In the hole doping region in Fig.~\ref{cap:mc}(a), the Hartree approach
underestimates the value of $m_{\textrm{c}}$ whereas the simple unscreened
result overestimates it. This can be attributed to the fact that the
Hartree theory used here is reliable only if the gap is small compared
to $t_{\perp}$. In the experimental realization of Ref.~\onlinecite{CNM+06},
$n_{0}>0$ and, therefore, the theory works well for a wide range
of electron doping $n>0$, whereas even a modest overall hole doping
$n<0$ corresponds to a significant electrostatic difference between
the two graphene layers. In this case, the unscreened theory overestimates
the gap whereas the Hartree calculation underestimates it. 

\begin{figure}
\begin{centering}
\includegraphics[clip,width=0.9\columnwidth]{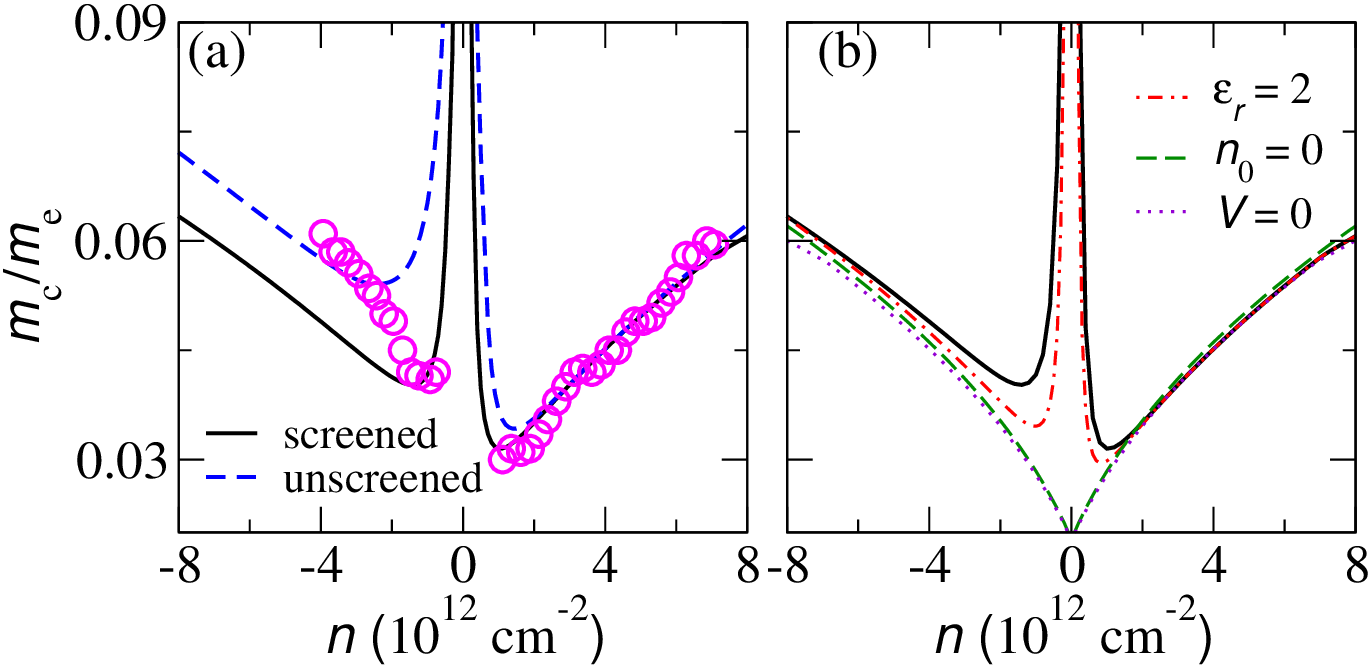} 
\par\end{centering}

\caption{\label{cap:mc}(Color online) Cyclotron mass vs $n$, normalized to
the free electron mass, $m_{\textrm{e}}$. (a)~Solid lines are the
result of the self-consistent procedure and the dashed lines correspond
to the unscreened case; $t\simeq3\,\mbox{eV}$, $t_{\perp}\simeq0.22\,\mbox{eV}$,
$\varepsilon_{r}=1$, and $n_{0}=1.8\times10^{12}\,\mbox{cm}^{-2}$.
Circles are experimental data from Ref.~\onlinecite{CNM+06}. (b)~The
screened result in~(a) is compared with the result for $\varepsilon_{r}=2$,
the case without chemical doping ($n_{0}=0$), and the case where
the external field is zero ($V=0$).}

\end{figure}

In Fig.~\ref{cap:mc}(b) we compare our best fit to the cyclotron
mass (full line) with results obtained for different parameter values.
The dashed-dotted lines stand for $m_{\textrm{c}}$ obtained with
$\varepsilon_{r}=2$ in Eq.~\eqref{eq:VnSgeim}. As can be seen clearly,
the $n>0$ result is not substantially affected, while for $n<0$
the theory description of $m_{\textrm{c}}$ worsens. This is due to
the reduction of the gap when $\varepsilon_{r}$ is increased {[}see
left inset in Fig.~\ref{fig:screening}(d){]}. The dashed lines in
Fig.~\ref{cap:mc}(b) are obtained with $n_{0}=0$, where the zero
gap occurs at the neutrality point. The dotted lines are the result
for $E_{ext}=0=V$, i.e., zero gap at every density value. Note that
these two results, $n_{0}=0$ and $V=0$, show an electron-hole symmetric
$m_{\textrm{c}}$, contradicting the experimental result. It may then
be said that the electron-hole asymmetry observed in $m_{\textrm{c}}$
is a clear indication of the presence of a finite gap in the spectrum.
It will be shown in Sec.~\ref{subsec:BilayerMFEmcEHasym} that, if
we ignore the gap, this electron-hole asymmetry cannot be described
by taken into account $t'$, $\gamma_{4}$ or $\Delta$.

\subsubsection{Cyclotron mass in continuum models}

\label{subsec:BilayerMFEmcCont}

Here we compare our results for the cyclotron mass, which has been
obtained with expressions shown in Appendix~\ref{secap:Bilayermc},
with the results of continuum models.

Within the 4-band continuum model given by Eq.~\eqref{eq:HKbl},
where the dispersion is just the full tight-binding result {[}Eq.~\eqref{eq:Ekbias}{]}
with the substitution $\epsilon_{\mathbf{k}}\to v_{\textrm{F}}p$,
we can easily derive the following analytical expression for $m_{\textrm{c}}$,\begin{equation}
m_{\textrm{c}}=\frac{E_{\textrm{F}}}{v_{\textrm{F}}^{2}}\left[1+\frac{V^{2}+t_{\perp}^{2}}{2\sqrt{E_{\textrm{F}}^{2}(V^{2}+t_{\perp}^{2})-t_{\perp}^{2}V^{2}/4}}\right].\label{eq:mc4bm}\end{equation}
In Fig.~\ref{fig:mccontEHasym}(a) the dashed line is the result
of Eq.~\eqref{eq:mc4bm}, where $V$ has been computed self-consistently
using Eq.~\eqref{eq:VnSgeim} and the 4-band continuum approximation
discussed in Sec.~\ref{subsubsec:GrapheneBEPscreenCont}. As expected,
the agreement with the full tight-binding result (shown as a full
line) is excellent for the considered densities. Note that there is
an extra solution given by $\tilde{m}_{\textrm{c}}v_{\textrm{F}}^{2}=E_{\textrm{F}}[1-(V^{2}+t_{\perp}^{2})/\sqrt{4E_{\textrm{F}}^{2}(V^{2}+t_{\perp}^{2})-t_{\perp}^{2}V^{2}}]$,
valid when $|E_{\textrm{F}}|<V/2$ or $|E_{\textrm{F}}|>\sqrt{V^{2}/4+t_{\perp}^{2}}$,
which corresponds to the extra orbit appearing when $E_{\textrm{F}}$
falls in the Mexican-hat region, or above the bottom of high energy
bands. We can estimate the densities for which these two regions start
playing a role: using $V\sim0.1t_{\perp}\sim0.01t$ in the Mexican
hat region (valid for $n_{0}\lesssim2\times10^{12}\,\mbox{cm}^{-2}$)
we get $n\lesssim10^{11}\,\mbox{cm}^{-2}$; setting $V\sim t_{\perp}\sim0.1t$
around the bottom of high energy bands we get $n\gtrsim10^{13}\,\mbox{cm}^{-2}$.
These two density values are outside the range of experimentally realized
densities {[}see Fig.~\ref{cap:mc}(a){]}.

\subsubsection{Effect of electron-hole asymmetry}

\label{subsec:BilayerMFEmcEHasym}

In Sec.~\ref{subsubsec:GrapheneBEPscreenEHasym} the effect of electron-hole
symmetry breaking parameters -- namely, $t'$, $\gamma_{4}$, and
$\Delta$ -- has been studied regarding the self-consistent description
of the gap. Here we extend the analysis to the cyclotron mass, restricting
ourselves to the biased BLG device shown in the left panel of Fig.~\ref{fig:device}(a).
Results have been obtained within the 4-band model. As all cases have
cylindrical symmetry around $K$ and $K'$, the cyclotron mass may
be written as $m_{\textrm{c}}=p_{\textrm{F}}/(\partial E_{\textrm{F}}/\partial p_{\textrm{F}})$.

\begin{figure}
\noindent \begin{centering}
\includegraphics[width=0.9\columnwidth]{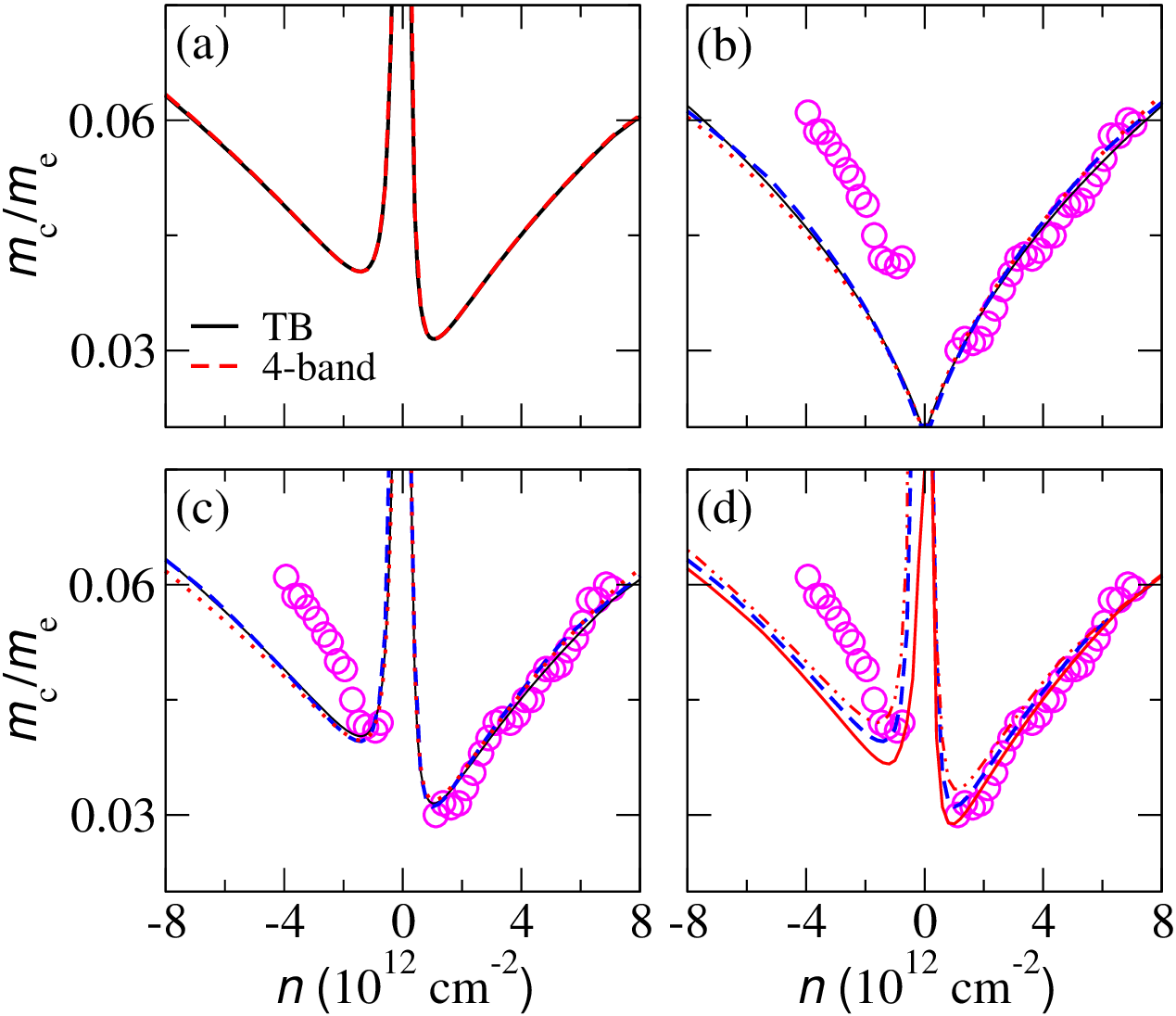}
\par\end{centering}

\caption{\label{fig:mccontEHasym}(Color online) Cyclotron mass vs $n$, normalized
to the free electron mass, $m_{\textrm{e}}$. (a)~Comparison between
full tight-binding (TB) and 4-band approximation for $t_{\perp}\simeq0.22\,\mbox{eV}$
and $n_{0}=1.8\times10^{12}\,\mbox{cm}^{-2}$. (b)-(c)~Effect of
finite $t'$ and $\gamma_{4}$ for $n_{0}=0$ and $n_{0}=1.8\times10^{12}\,\mbox{cm}^{-2}$,
respectively: dotted line is for $t'\simeq0.3\,\mbox{eV}$ and $t_{\perp}\simeq0.22\,\mbox{eV}$;
dashed line is for $\gamma_{4}\simeq0.12\,\mbox{eV}$ and $t_{\perp}\simeq0.19\,\mbox{eV}$;
full thin line is for $t'=\gamma_{4}=0$ and $t_{\perp}=0.22\,\mbox{eV}$.
(d)~Effect of $\Delta$ for $\gamma_{4}\simeq0.12\,\mbox{eV}$ and
$t_{\perp}\simeq0.19\,\mbox{eV}$: full line for $\Delta\simeq0.03\,\mbox{eV}$;
dotted-dashed line for $\Delta\simeq-0.03\,\mbox{eV}$; dashed line
for $\Delta=0$. Circles are experimental data from Ref.~\onlinecite{CNM+06}.
We have used $t\simeq3\,\mbox{eV}$ and $\varepsilon_{r}=1$.}

\end{figure}

In Fig.~\ref{fig:mccontEHasym}(b) we show the $m_{\textrm{c}}$
result for finite $t'$ (dotted red line) and finite $\gamma_{4}$
(dashed blue line), keeping $n_{0}=0$ (absence of electron-hole asymmetry
due to chemical doping). The thin full line is the result obtained
for $t'=\gamma_{4}=0$ in Sec.~\ref{subsec:BilayerMFEmcCompExp},
and circles are experimental data from Ref.~\onlinecite{CNM+06}.
The $n>0$ region, where the smaller gaps are realized experimentally,
is still well fitted if we choose $t_{\perp}\simeq0.22\,\mbox{eV}$
with $t'\simeq0.3\,\mbox{eV}$ or $t_{\perp}\simeq0.19\,\mbox{eV}$
with $\gamma_{4}=0.12\,\mbox{eV}$ (we use $t\simeq3\,\mbox{eV}$).
However, it is clear that neither of these results can account for
the electron-hole asymmetry observed experimentally. In fact, a closer
look reveals that the $m_{\textrm{c}}$ for finite $t'$ have the
opposite trend, being smaller than the $t'=0$ result for $n<0$ and
larger for $n>0$, as would be expected by inspection of the energy
bands in Fig.~\ref{fig:ehAsym}(a). Such an opposite trend should
also be seen for finite $\gamma_{4}$, although the effect is not
as large as expected from the considerable distortion of the energy
bands shown in Fig.\ref{fig:ehAsym}(b). This attenuation can be understood
as the result of fixing the carrier density $n$ and not the Fermi
energy $E_{\textrm{F}}$: changing $\gamma_{4}$ (or $t'$) for a
given $n$ leads to a different $E_{\textrm{F}}$, and the new $E_{\textrm{F}}$
is such that it counteracts the expected effect of $\gamma_{4}$ (or
$t'$) in $m_{\textrm{c}}$. Fig.~\ref{fig:mccontEHasym}(c) shows
the same as \ref{fig:mccontEHasym}(b) for $n_{0}=1.8\times10^{12}\,\mbox{cm}^{-2}$.
The effect of the on-site energy $\Delta$ is shown in Fig.~\ref{fig:mccontEHasym}(d)
for fixed $\gamma_{4}\simeq0.12\,\mbox{eV}$, $t_{\perp}\simeq0.19\,\mbox{eV}$
and $n_{0}=1.8\times10^{12}\,\mbox{cm}^{-2}$. The result for $\Delta=0$
(dashed line) is shown along with the result for $\Delta\simeq0.03\,\mbox{eV}$
(full line) and $\Delta\simeq-0.03\,\mbox{eV}$ (dotted-dashed line).
It is clear that the effect of $t'$, $\gamma_{4}$ and $\Delta$
on the cyclotron mass can be neglected.

%

\subsection{Cyclotron resonance}

\label{subsec:BilayerMFEllCR}

The effect of a perpendicular magnetic field can be studied within
the continuum approximation through minimal coupling $\mathbf{p}\to\mathbf{p}-e\mathbf{A}$.\citep{MF06}
The case of biased BLG has been studied both within the 4-band {[}Eq.~\eqref{eq:HKbl}{]}
and 2-band {[}Eq.~\eqref{eq:Heff}{]} continuum models in Refs.~\onlinecite{McC06,GNP06,PPV07,MS08}.
Here we use the same approach to study the cyclotron resonance (i.e.
the Landau level transition energies) with the extra ingredient that
the parameter $V$ depends on the filling factor, as discussed in
Sec.~\ref{subsec:screen}.

In the 4-band model standard manipulations\citep{MF06,GNP06,PPV07,NHI08}
lead to the unbiased BLG Landau level spectrum \begin{equation}
E_{n}^{\pm\pm}=\pm\sqrt{(1+2n)\frac{\gamma^{2}}{2}+\frac{t_{\perp}^{2}}{2}\pm\sqrt{(\gamma^{2}+t_{\perp}^{2})^{2}/4+n\gamma^{2}t_{\perp}^{2}}},\label{eq:LLbilayer}\end{equation}
where $\gamma=\sqrt{2}v_{\textrm{F}}\hbar/l_{B}$, with $l_{B}=\sqrt{\hbar/|e|B}$
for the magnetic length. Non-zero ($n\geq1$) eigenenergies are fourfold
degenerate due to valley and spin degeneracy, while zero energy Landau
levels have eightfold degeneracy, since there are two zero energy
Landau states ($n=-1,0$) per valley per spin. The 2-band model result
$E_{n}^{\pm}\approx\pm\gamma^{2}t_{\perp}^{-1}\sqrt{n(n+1)}$ is easily
recovered from Eq.~(\ref{eq:LLbilayer}) for $\gamma\ll t_{\perp}$,
being valid for magnetic fields up to $B\approx1\,\textrm{T}$.\citep{MF06}

\begin{figure}
\begin{centering}
\includegraphics[width=0.75\columnwidth]{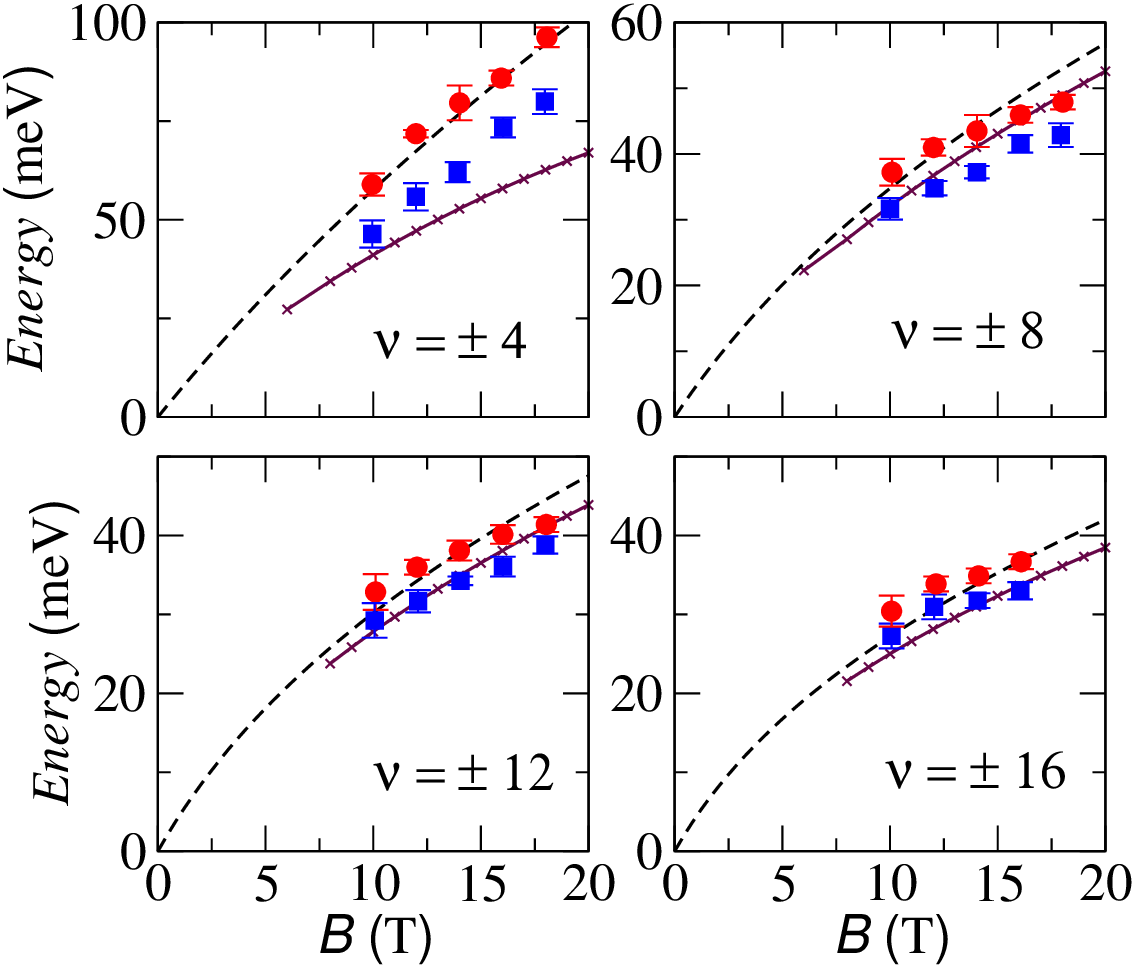}
\par\end{centering}

\caption{\label{fig:cycloRes}(Color online) Landau level transition energies
vs magnetic field for the given filling factors. The dashed line is
the unbiased BLG result {[}Eq.~\eqref{eq:LLbilayer}{]} and the line
with crosses is the biased BLG result (see text). We used $t=3.5\,\mbox{eV}$
and $t_{\perp}=0.1t$. Filled symbols are experimental data from Ref.~\onlinecite{HJT+08}:
circles for electrons and squares for holes. }

\end{figure}

The Landau level transition energies in BLG have been recently obtained
through cyclotron resonance measurements.\citep{HJT+08} The data
was found to deviate from what would be expected through Eq.~\eqref{eq:LLbilayer},
especially for larger filling factors. It should be noted, however,
that in order to keep a constant filling factor and vary the magnetic
field, as done in Ref.~\onlinecite{HJT+08}, the back gate voltage
$V_{g}$ has to be tuned to compensate for the variation of Landau
level degeneracy. As we have seen previously, tuning $V_{g}$ is equivalent
to change $V$ -- the electrostatic energy difference between layers
-- which means that Eq.~\eqref{eq:LLbilayer} is no longer valid,
as recently shown within the 4-band continuum model.\citep{PPV07}
To have an estimate for the effect of the back gate voltage on the
Landau level spacing we have computed Landau level energy differences
taking into account the variation of $V$ with carrier density $n$.
We have used the unscreened result given by Eq.~\eqref{eq:VnUnSgeim},
with $n_{0}=0$ and $\varepsilon_{r}=1$. Within this approximation
we can easily write $V$ in terms of the filling factor $\nu$ and
magnetic field $B$ as $V=\nu Be^{2}d/(2\varepsilon_{0}\phi_{0})\approx7.4\times10^{-4}\nu B$,
with $B$ in Tesla in the last step. Thus, for fixed filling factor,
$V$ varies linearly with $B$. Note that the comparison between this
unscreened treatment of the biased BLG and the unbiased result in
Eq.~\eqref{eq:LLbilayer} should give lower and upper limits for
the effect of the perpendicular external field in the cyclotron frequency.
In Fig.~\ref{fig:cycloRes} we show the obtained Landau transition
energies vs magnetic field for the given filling factors. The dashed
lines represent the unbiased BLG result, as given by Eq.~\eqref{eq:LLbilayer}.
The lines with crosses are the results for the unscreened biased BLG,
and filled symbols are experimental data from Ref.~\onlinecite{HJT+08}:
circles for $\nu>0$ and squares for $\nu<0$. We have used $t=3.5\,\mbox{eV}$
and $t_{\perp}=0.1t$, consistent with Ref.~\onlinecite{HJT+08}.
As can be seen from Fig.~\ref{fig:cycloRes}, the back gate induced
electric field gives rise to sizable effects already for magnetic
fields and filling factors realized in experiments. Except at $\nu=\pm8$,
the result of Eq.~\eqref{eq:LLbilayer} for the unbiased BLG and
the unscreened biased BLG result effectively provide upper and lower
limits to the experimental data. The observed electron-hole asymmetry
could then be interpreted as due to an asymmetry in $V$ vs $n$:
larger $V$, and therefore larger gap, for $n<0$; smaller $V$ and
gap for $n>0$, which would make the result more close to the unbiased
case. It should be noted that in such a case we would expect the neutrality
point to occur for $V_{g}<0$, as is the case of the NH$_{3}$ doped
BLG studied before. For the system reported in Ref.~\onlinecite{HJT+08},
however, the opposite seems to be happening, as indicated by the Hall
resistivity. A neutrality point for $V_{g}>0$ is, in fact, the more
usual effect of H$_{2}$O molecules adsorbed on graphene samples.\citep{SGM+07}
As a final remark regarding the results presented in Fig.~\ref{fig:cycloRes},
we note that the experimental data trend, which makes Eq.~\eqref{eq:LLbilayer}
a poor fit at $|\nu|\geq8$, is still not accounted for in the biased
BLG result. An alternative approach is the inclusion of the screening
correction, which should go beyond Eq.~\eqref{eq:Dnk} including
the magnetic field effect. It has been reported recently that Dirac
liquid renormalization may also be contributing to the observed trend.\citep{KCN08}

%

\section{Conclusions}

\label{sec:conclusions}

We have studied the electronic behavior of bilayer graphene in the
presence of a perpendicular electric field -- \emph{biased bilayer}
-- using the minimal tight-binding model that describes the system.
The effect of the perpendicular electric field has been included through
a parallel plate capacitor model, with screening correction at the
Hartree level. We have compared the full tight-binding description
with its 4-band and 2-band continuum approximations, and found that
the 4-band model is always a suitable approximation for the conditions
realized in experiments. Also, we have studied the effect of electron-hole
asymmetry terms and found that they have only a small effect on the
electronic properties addressed here. The model has been applied to
real biased bilayer devices, either made out of SiC\citep{OBS+06}
or exfoliated graphene.\citep{CNM+06,OHL+07} The good agreement with
experimental results -- namely, for the electrostatic energy difference
between layers obtained through ARPES\citep{OBS+06} and for the Shubnikov-de
Haas cyclotron mass\citep{CNM+06} -- clearly indicates that the model
is capturing the key ingredients, and that a finite gap is effectively
being controlled externally. Analysis of recent experimental results
regarding the electrical noise\citep{LA08} and cyclotron resonance\citep{HJT+08}
further suggests that the model can be seen as a good starting point
to understand the electronic properties of graphene bilayer. 


\subsection*{Acknowledgments}

E.V.C., N.M.R.P., and J.M.B.L.S. acknowledge financial support from
POCI 2010 via project PTDC/FIS/64404/2006. A.H.C.N. acknowledges the
partial support of the U.S. Department of Energy under grant No. DE-FG02-
08ER46512.


\appendix

%

\section{Asymmetry between layers\label{secap:Dn}}

In order to write Eq.~\eqref{eq:Dnk} as an energy integral, we start
by introducing the SLG density of states per spin per unit cell defined
for the conduction band as \begin{equation}
\rho(\epsilon)=\frac{1}{N_{\textrm{c}}}\sum_{\mathbf{k}}\delta(\epsilon-t\vert s_{\mathbf{k}}\vert),\label{Eq_rho}\end{equation}
with $s_{\mathbf{k}}$ as in Eq.~\eqref{eq:HkBilayer}. The momentum
sum in Eq.~(\ref{Eq_rho}) can be written as an integral by letting
$N_{\textrm{c}}\rightarrow\infty$. The integral can be performed
and written in terms of complete elliptic integrals of the first kind.\citep{NGPrmp}

With the definition of $\rho(\epsilon)$ in Eq.~\eqref{Eq_rho} the
charge imbalance between layers in Eq.~\eqref{eq:Dnk} can be written
as $\Delta n=\Delta n_{1/2}+\Delta\tilde{n}$, where the charge imbalance
at half-filling $\Delta n_{1/2}$ is given by\begin{equation}
\Delta n_{1/2}=\frac{2}{A_{\hexagon}}\sum_{l=\pm}\int_{0}^{3t}\textrm{d}\epsilon\,\rho(\epsilon)\mathcal{I}_{-l}(\epsilon),\label{eq:DnHalf}\end{equation}
and the fluctuation $\Delta\tilde{n}$ with respect to the half-filled
case is given by

\begin{equation}
\Delta\tilde{n}=\frac{2}{A_{\hexagon}}\begin{cases}
\sum_{l=\pm}\int_{\epsilon_{1}}^{\epsilon_{2}}\textrm{d}\epsilon\,\rho(\epsilon)\mathcal{I}_{+l}(\epsilon)\,, & n>0\\
-\sum_{l=\pm}\int_{\epsilon_{1}}^{\epsilon_{2}}\textrm{d}\epsilon\,\rho(\epsilon)\mathcal{I}_{-l}(\epsilon)\,, & n<0\end{cases},\label{eq:DnTilde}\end{equation}
where $n$ is the carrier density with respect to half-filling. The
integral kernel in Eqs.~\eqref{eq:DnHalf} and~\eqref{eq:DnTilde}
reads\begin{multline}
\mathcal{I}_{jl}(\epsilon)=\\
\frac{[\epsilon^{2}+\mathcal{K}_{-}^{jl}(\epsilon)](\epsilon^{2}-\mathcal{K}_{+}^{jl})^{2}-(\epsilon^{2}+\mathcal{K}_{+}^{jl})t_{\perp}^{2}\mathcal{K}_{-}^{jl}(\epsilon)}{[\epsilon^{2}+\mathcal{K}_{-}^{jl}(\epsilon)][\epsilon^{2}-\mathcal{K}_{+}^{jl}(\epsilon)]^{2}+[\epsilon^{2}+\mathcal{K}_{+}^{jl}(\epsilon)]t_{\perp}^{2}\mathcal{K}_{-}^{jl}(\epsilon)},\label{eq:Iener}\end{multline}
where $\mathcal{K}_{\pm}^{jl}(\epsilon)=[V/2\pm E^{jl}(\epsilon)]^{2}$,
with $E^{jl}(\epsilon)$ given by Eq.~\eqref{eq:Ekbias} with the
substitution $\epsilon_{\mathbf{k}}\rightarrow\epsilon$. The limits
of integration in Eq.~(\ref{eq:DnTilde}) depend on the band label
$l$ and $E_{\textrm{F}}$ as follows: with $l=-$ we have $\epsilon_{1}=\epsilon^{-}$
and $\epsilon_{2}=\epsilon^{+}$ for $E_{\textrm{F}}^{2}<V^{2}/4$,
while for $E_{\textrm{F}}^{2}>V^{2}/4$ we have $\epsilon_{1}=0$
and $\epsilon_{2}=\epsilon^{+}$; with $l=+$ we only have contribution
for $E_{\textrm{F}}^{2}>t_{\perp}^{2}+V^{2}/4$, and the limits are
$\epsilon_{1}=0$ and $\epsilon_{2}=\epsilon^{-}$. We use the notation
$\epsilon^{\pm}=[E_{\textrm{F}}^{2}+V^{2}/4\pm\sqrt{E_{\textrm{F}}^{2}(V^{2}+t_{\perp}^{2})-t_{\perp}^{2}V^{2}/4}]^{\frac{1}{2}}.$

%

\section{Bilayer DOS}

\label{secap:BilayerDOS}

The DOS per unit cell of BLG, either biased or unbiased, is defined
as\begin{equation}
\rho_{2}(E)=\frac{2}{N_{\textrm{c}}}\sum_{\mathbf{k}}[\delta(E-E_{\mathbf{k}}^{\pm-})+\delta(E-E_{\mathbf{k}}^{\pm+})],\label{eq:DOS2def}\end{equation}
where $E_{\mathbf{k}}^{\pm\pm}$ is given by Eq.~(\ref{eq:Ekbias}).
Equation~\eqref{eq:DOS2def} can be written as a sum of two contributions,
$\rho_{2}(E)=\sum_{l=\pm}\rho_{2}^{l}(E)$, where the label $l=\pm$
stands for contributions coming from bands $E_{\mathbf{k}}^{\pm l}$.
The analytical expressions for each contribution are\begin{widetext}
\begin{equation}
\rho_{2}^{-}(E)=\frac{4}{t^{2}\pi^{2}}\begin{cases}
\psi^{--}(E)\frac{\chi^{-}(E)}{\sqrt{F[\chi^{-}(E)/t]}}\mathbf{K}\left(\frac{4\chi^{-}(E)/t}{F[\chi^{-}(E)/t]}\right), & \begin{cases}
\Delta_{g}/2<|E|<V/2\wedge\alpha\leq t^{2}\\
\hspace{1cm}\vee\\
E^{+-}(t)<|E|<V/2\wedge\alpha>t^{2}\end{cases}\\
\hspace{2cm}+\\
\psi^{-+}(E)\frac{\chi^{+}(E)}{\sqrt{F[\chi^{+}(E)/t]}}\mathbf{K}\left(\frac{4\chi^{+}(E)/t}{F[\chi^{+}(E)/t]}\right), & \Delta_{g}/2<|E|<E^{+-}(t)\wedge\alpha<t^{2}\\
\\\psi^{--}(E)\frac{\chi^{-}(E)}{\sqrt{4\chi^{-}(E)/t}}\mathbf{K}\left(\frac{F[\chi^{-}(E)/t]}{4\chi^{-}(E)/t}\right), & \Delta_{g}/2<|E|<E^{+-}(t)\wedge\alpha>t^{2}\\
\hspace{2cm}+\\
\psi^{-+}(E)\frac{\chi^{+}(E)}{\sqrt{4\chi^{+}(E)/t}}\mathbf{K}\left(\frac{F[\chi^{+}(E)/t]}{4\chi^{+}(E)/t}\right), & \begin{cases}
E^{+-}(t)<|E|<E^{+-}(3t)\wedge\alpha<t^{2}\\
\hspace{1cm}\vee\\
\Delta_{g}/2<|E|<E^{+-}(3t)\wedge t^{2}\leq\alpha<9t^{2}\end{cases}\end{cases}\label{eq:DOS2m}\end{equation}

and\begin{equation}
\rho_{2}^{+}(E)=\frac{4}{t^{2}\pi^{2}}\begin{cases}
\psi^{+-}(E)\frac{\chi^{-}(E)}{\sqrt{F[\chi^{-}(E)/t]}}\mathbf{K}\left(\frac{4\chi^{-}(E)/t}{F[\chi^{-}(E)/t]}\right), & \sqrt{t_{\perp}^{2}+V^{2}/4}<|E|<E^{++}(t)\\
\psi^{+-}(E)\frac{\chi^{-}(E)}{\sqrt{4\chi^{-}(E)/t}}\mathbf{K}\left(\frac{F[\chi^{-}(E)/t]}{4\chi^{-}(E)/t}\right), & E^{++}(t)<|E|<E^{++}(3t)\end{cases},\label{eq:DOS2p}\end{equation}
 with $\psi^{\pm l}(E)$ given by \begin{equation}
\psi^{\pm l}(E)=\frac{\sqrt{t_{\perp}^{4}/4+(t_{\perp}^{2}+V^{2})\chi^{l}(E)^{2}}\sqrt{\chi^{l}(E)^{2}+t_{\perp}^{2}/2+V^{2}/4\pm\sqrt{t_{\perp}^{4}/4+(t_{\perp}^{2}+V^{2})\chi^{l}(E)}}}{\chi^{l}(E)\Big|\sqrt{t_{\perp}^{4}/4+(t_{\perp}^{2}+V^{2})\chi^{l}(E)^{2}}\pm(t_{\perp}^{2}+V^{2})/2\Big|}\label{eq:psi}\end{equation}
\end{widetext} and $\chi^{\pm}(E)$ as in the right-hand side of
Eq.~\eqref{eq:kpmLim} with $E_{\mathrm{F}}\rightarrow E$. We use
$F(x)=(1+x)^{2}-(x^{2}-1)^{2}/4$ and $\mathbf{K}(m)$ for the complete
elliptic integral of the first kind, and $E^{\pm\pm}(x)$ is given
by Eq.~\eqref{eq:Ekbias} with the substitution $\epsilon_{\mathbf{k}}\rightarrow x$
and $\alpha=(V^{4}/4+t_{\perp}^{2}V^{2}/2)/(V^{2}+t_{\perp}^{2})$.

%

\section{Cyclotron mass}

\label{secap:Bilayermc}

Based on the full tight-binding band structure $E_{\mathbf{k}}^{\pm\pm}$
given in Eq.~\eqref{eq:Ekbias}, it is possible to derive general
expressions for the cyclotron mass in Eq.~\eqref{eq:mcdef}. The
key observation is that the area of a closed orbit at the Fermi level
$A(E_{\textrm{F}})$ may be written as $A(E_{\textrm{F}})\propto\sideset{}{'}\sum_{\mathbf{k}}\Delta_{\mathbf{k}}$,
where the prime means summation over all $\mathbf{k}$'s inside the
orbit, and $\Delta_{\mathbf{k}}=(2\pi)^{2}/(N_{\textrm{c}}A_{\hexagon})$
is the area per $k$-point in the first BZ. The cyclotron mass may
then be written as $m_{\textrm{c}}\propto\partial_{E_{\textrm{F}}}\sum_{i}\int_{\mathcal{E}_{i}}^{E_{\textrm{F}}}\mbox{d}E\,\sum_{\mathbf{k}}\delta(E-E_{\mathbf{k}}^{\mu\nu})\Theta(\epsilon^{\pm}-\epsilon_{\mathbf{k}})$,
where $\epsilon_{\mathbf{k}}$ is the SLG dispersion and $\epsilon^{\pm}$
is given by Eq.~\eqref{eq:Ekbias}. The integration limits $\mathcal{E}_{i}$
and the choice between the two possibilities $\epsilon^{\pm}$ depend
on the particular band and on the position of the Fermi level. Skipping
the details of the derivation, what is worth noting is that, due to
the sum of delta functions in the previous expression for $m_{\textrm{c}}$,
the result has a mathematical structure similar to the derived expressions
for the DOS of BLG (see Appendix~\ref{secap:BilayerDOS}). The cyclotron
effective mass of the biased BLG for the relevant parameter range
$V\lesssim t_{\perp}\ll t$ and $|E_{\textrm{F}}|\lesssim t$ is then
given by\begin{widetext} \begin{equation}
m_{\textrm{c}}(E_{\textrm{F}})=\frac{\hbar^{2}}{A_{\hexagon}t^{2}}\frac{2}{\pi}\begin{cases}
-\psi^{--}(E_{\textrm{F}})\frac{\chi^{-}(E_{\textrm{F}})}{\sqrt{F[\chi^{-}(E_{\textrm{F}})/t]}}\mathbf{K}\left(\frac{4\chi^{-}(E_{\textrm{F}})/t}{F[\chi^{-}(E_{\textrm{F}})/t]}\right), & \Delta_{g}/2<|E_{\textrm{F}}|<V/2\\
\psi^{-+}(E_{\textrm{F}})\frac{\chi^{+}(E_{\textrm{F}})}{\sqrt{F[\chi^{+}(E_{\textrm{F}})/t]}}\mathbf{K}\left(\frac{4\chi^{+}(E_{\textrm{F}})/t}{F[\chi^{+}(E_{\textrm{F}})/t]}\right), & \Delta_{g}/2<|E_{\textrm{F}}|\lesssim t\\
\psi^{+-}(E_{\textrm{F}})\frac{\chi^{-}(E_{\textrm{F}})}{\sqrt{F[\chi^{-}(E_{\textrm{F}})/t]}}\mathbf{K}\left(\frac{4\chi^{-}(E_{\textrm{F}})/t}{F[\chi^{-}(E_{\textrm{F}})/t]}\right), & \sqrt{t_{\perp}^{2}+V^{2}/4}<|E_{\textrm{F}}|\lesssim t\end{cases}.\label{eq:mcG}\end{equation}
\end{widetext}

\bibliographystyle{apsrev}

\bibliographystyle{apsrev}

\begin{thebibliography}{88}
\bibitem[{foo({\natexlab{b}})}]{foot2}
\bibinfo{note}{Although the buffer layer concept seems to work well for the
  Si-face, its validity for the C-face is still under debate. See
  Refs.~\onlinecite{HFM+07,ESS+08,SR09}.}
\bibitem[{foo({\natexlab{c}})}]{foot3}
\bibinfo{note}{Interface states seem to be a robust concept, even though the
  structure of the buffer layer at the Si-face might be more complex than
  previously anticipated. See Ref.~\onlinecite{HOF+08}.}
\end{thebibliography}

\end{document}